\documentclass[letterpaper, 10 pt, conference]{ieeeconf}  

\IEEEoverridecommandlockouts                          \usepackage[utf8]{inputenc}
\usepackage{xcolor}
\usepackage{dirtytalk}
\usepackage{graphicx}
\usepackage{graphics}
\usepackage{graphicx}
\usepackage{subcaption}
\usepackage{amsmath}
\usepackage[version=4]{mhchem}
\usepackage{siunitx}
\usepackage{longtable,tabularx}
\usepackage{algorithm}
\usepackage{algpseudocode}
\usepackage{amsmath}
\usepackage{amssymb}
\usepackage{proof}
\usepackage{epsfig}

\newtheorem{problem}{Problem}

\renewcommand{\baselinestretch}{0.98}

\title{\LARGE \bf
Far-Field Minimum-Fuel Spacecraft Rendezvous using Koopman Operator and $\ell_2/\ell_1$ Optimization
}

\author{Vrushabh Zinage$^{1}$ \and Efstathios Bakolas$^{2}$
\thanks{$^{1}$Vrushabh Zinage is a graduate student at the Department of Aerospace Engineering and Engineering Mechanics, University of Texas at Austin, Austin, Texas 78712, 
        {\tt\small vrushabh.zinage@gmail.com }}%
\thanks{$^{2}$Efstathios Bakolas is an Associate Professor in the Department of Aerospace Engineering and Engineering Mechanics, University of Texas at Austin, Austin, Texas 78712,       {\tt\small bakolas@austin.utexas.edu}}%
}

\begin{document}

\bibliographystyle{IEEEtran} 

\maketitle
\thispagestyle{empty}
\pagestyle{empty}

\begin{abstract}
We propose a method to compute approximate solutions to the minimum-fuel far-field rendezvous problem for thrust-vectoring spacecraft. 
It is well-known that the use of linearized spacecraft rendezvous equations may not give sufficiently accurate results for far-field rendezvous. In particular, as the distance between the active and the target spacecraft becomes significantly greater than the distance between the target spacecraft and the center of gravity of the planet, the accuracy of linearization-based control design approaches may decline substantially. In this paper, we use a nonlinear state space model which corresponds to more accurate description of dynamics than linearized models but at the same time poses the known challenges of nonlinear control design. To overcome these challenges, we utilize a Koopman operator based approach with which the nonlinear spacecraft rendezvous dynamics is lifted into a higher dimensional space over which the nonlinear dynamics can be approximated by a linear system which is more suitable for control design purposes than the original nonlinear model. An Iteratively Recursive Least Squares (IRLS) algorithm from compressive sensing is then used to solve the minimum fuel control problem based on the lifted linear system. 
%
Numerical simulations are performed to show the efficacy of the proposed Koopman operator based approach.

\end{abstract}

\section{Introduction}
We propose a Koopman operator based method for the computation of control inputs that correspond to approximate solutions to the minimum-fuel far-field rendezvous problem for thrust-vectoring spacecraft. In a typical rendezvous problem, the relative motion of the active chaser spacecraft with respect to a target spacecraft in a circular or elliptical orbit can be described in terms of a system of autonomous nonlinear differential equations. The control design in such problems is based, however, on linearized equations of motion such as the Hill–Clohessy–Wiltshire (H–C–W) equations, which correspond to a time-invariant system of equations, or the Tschauner–Hempel (T–H) equations, which correspond to a periodic linear system. 
%
%
These widely used linearized models are rarely effective to describe the relative motion for far-field rendezvous \cite{walsh2017general_far_rendezvous}.
Therefore, linearization-based control design techniques cannot guarantee the desired accuracy in far-field rendezvous problems. The Koopman operator approach utilized herein allows one to account for the nonlinearities of the dynamics of the spacecraft rendezvous problem while at the same time linear control design techniques are still applicable. The key idea of the Koopman operator is that the nonlinear dynamics of the rendezvous problem can be approximated by a higher dimensional linear state space model based on which we can compute approximate solutions to the minimum-fuel rendezvous problem for a thrust vectoring spacecraft. The proposed control algorithms rely on tools from compressive sensing \cite{b:Foucart2013} and in particular for $\ell_2/\ell_1$ optimization and the Iteratively Reweighted Least Squares algorithm \cite{sparse_bakolas2019computation,bakolas2018solution,block_wang2013recovery_sparse}.


\textit{Literature review:} A rendezvous mission is usually divided into far-field rendezvous, near-field
rendezvous, and final approach. Various control approaches have been proposed for near-field and final approach rendezvous operations \cite{prussing1969optimal_conv2,carter1995linearized_conv3,arzelier2013new_conv4,leomanni2019sum,huber_karlgaard2006robust_huber,adpative_singla2006adaptive,neural_youmans1998neural,gao2009multi_h_infinity,yao2010flyaround_relative,luo2007optimal,lopez1995autonomous_artificial_1,breger2008safe}. Some of these include multi-objective robust $H_{\infty}$ control \cite{gao2009multi_h_infinity}, neural network approach \cite{neural_youmans1998neural}, adaptive control methods \cite{adpative_singla2006adaptive}, a Huber filter approach \cite{huber_karlgaard2006robust_huber} and artificial potential function approaches \cite{lopez1995autonomous_artificial_1}. However, most of these approaches use linearized rendezvous equations.
References \cite{ping_lu2013autonomous_cone_2,liu2013robust,ping_liu2014solving,p:acikmese2017} consider more general and challenging proximity operation problems under realistic constraints. While these methods are very robust, they are rarely effective for far-field rendezvous \cite{walsh2017general_far_rendezvous}. In addition, these methods \cite{huber_karlgaard2006robust_huber,adpative_singla2006adaptive,neural_youmans1998neural,gao2009multi_h_infinity,yao2010flyaround_relative,luo2007optimal} cannot be used for far-field rendezvous as these linearized equations give inaccurate results, are computationally expensive and do not guarantee any optimality in terms of fuel consumption.

Koopman operator is an infinite dimensional linear operator that describes the evolution of functions of states (referred to as observable functions or just observables). This operator allows one to ``convert'' a finite-dimensional nonlinear system into a linear system by lifting the state space of the former system to a higher dimensional state space over which it admits a linear, yet infinite-dimensional, state space model representation. This lifting approach can be traced back to earlier works of \cite{koopman1931hamiltonian_main_1,koopman1932dynamical_main_2}. However, in practical applications, a finite-dimensional approximation of the Koopman operator can provide a sufficiently accurate description of the evolution of a nonlinear dynamical systems. By applying linear control design techniques to the system on the \say{lifted} state space, one obtains indirectly a controller that can be applied to the original nonlinear system of interest~  \cite{brunton2016koopman_control_1,proctor2016dynamic_koopman_control_2,proctor2018generalizing_koopman_control_3,williams2016extending_koopman_control_4}. References \cite{brunton2016koopman_control_1,proctor2016dynamic_koopman_control_2,proctor2018generalizing_koopman_control_3,williams2016extending_koopman_control_4} propose extensions of the Koopman operator approach for control systems. 
The works of \cite{proctor2016dynamic_koopman_state_1,surana2016koopman_state_2,surana2016linear_koopman_state_3} use Koopman operator methods for state estimation and nonlinear system identification.
Recent studies on the computation of finite-dimensional approximations to the Koopman operator that lead to better approximations of nonlinear dynamics can be found in \cite{williams2015data_koopman_accuracy_1}. 
%
%
A systematic process to choose the observable functions that can best approximate the Koopman operator remains, however, an open research problem. Some recent efforts to address the latter problem based on a combination of machine learning and trial and error methods can be found in \cite{lusch2018deep_machine_learning, abraham2019active_trial}. In some cases, the choice of observable functions is system-specific \cite{mauroy2016linear_state_specfic}.

\textit{Main contributions:} In this paper, we use the Koopman operator to lift the nonlinear spacecraft rendezvous dynamics into a higher but finite-dimensional space over which it can be approximated by a linear system. An Iteratively Recursive Least Squares (IRLS) \cite{block_wang2013recovery_sparse} algorithm is then used to compute approximate solutions for control sequences that minimize the fuel consumption for far-field rendezvous of a thrust vectoring spacecraft. Through numerical simulations, it is observed that the Koopman based approach is able to steer the active spacecraft to the desired final states for both short-field and far-field rendezvous with higher accuracy than when the same controller is designed based on the linearized model for rendezvous. The superiority of the Koopman approach over the standard linearization-based approach is more significant in the case of far-field rendezvous, in which the latter often gives significantly large miss-target errors. To the best knowledge of the authors, this is the first paper which utilizes the Koopman operator for the minimum-fuel spacecraft rendezvous problem.

{\textit{Structure of the paper:}} The organization of the paper is as follows. In Section \ref{sec:state_space_model}, the continuous-time and discrete-time nonlinear state space models for spacecraft rendezvous are introduced. Koopman operator is reviewed in Section \ref{sec:koopman_operator}. Section \ref{sec:algorithms} introduces the proposed solution approach for the minimum fuel problem based on the IRLS algorithm for thrust vectoring spacecraft. Numerical simulations are presented in Section \ref{sec:numerical_simulation_results}, and Section \ref{sec:conclusions} presents concluding remarks.

\section{State space model and problem setup}\label{sec:state_space_model}
In this section, we briefly discuss the governing equations and introduce continuous-time and discrete-time state space models for spacecraft rendezvous. Then, we introduce the problem addressed in this paper.

\begin{figure}[h]
\centering
\includegraphics[width=8.5cm]{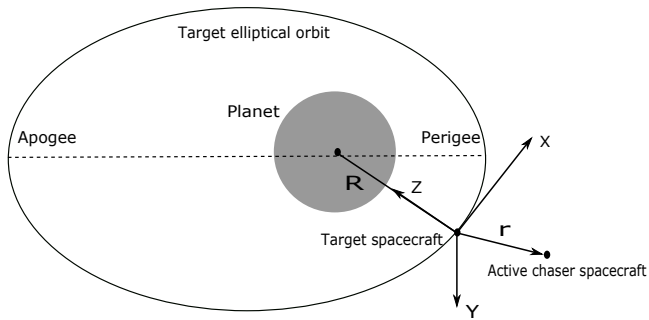}
\caption{Local Vertical Local Horizontal (LVLH) coordinate system for spacecraft rendezvous}
\label{fig:coorinate_frame}
\end{figure}
Assume that the target spacecraft is in an elliptical orbit with eccentricity $e$. Consider the Local-Vertical-Local-Horizontal coordinate system $X-Y-Z$ as shown in Fig. \ref{fig:coorinate_frame} where the origin is fixed at the center of mass of the target spacecraft, and the $Y$ axis is normal to the orbital plane $X-Z$. 
The relative motion of the active chaser spacecraft in the LVLH frame can be captured by the following nonlinear equation \cite{yamanaka2002new_dynamics_state_trans}:
\begin{align}
\frac{\mathrm{d}^{2} \boldsymbol{r}}{\mathrm{d} t^{2}}=-\mu\left(\frac{\boldsymbol{R}+\boldsymbol{r}}{|\boldsymbol{R}+\boldsymbol{r}|^{3}}-\frac{\boldsymbol{R}}{|\boldsymbol{R}|^{3}}\right)+\boldsymbol{u},
\label{eqn:nonlinear_space_rendezvous_governing}
\end{align}
where $\mu$ is the gravity constant, $\boldsymbol{u}$ is the control input (acceleration vector due to thrust forces on the active chaser spacecraft), $\boldsymbol{r}$ is the vector from the target spacecraft to the active chaser spacecraft and $\boldsymbol{R}$ is the relative position vector from the center of gravity of the planet to the target spacecraft.

\subsection{Continuous-time nonlinear model}

Using the notation $\boldsymbol{r}=\left[\begin{array}{lll}x & y & z\end{array}\right]^\mathrm{T},$ Eq. \eqref{eqn:nonlinear_space_rendezvous_governing} can be written as \cite{yamanaka2002new_dynamics_state_trans}
\begin{align}
\left[\begin{array}{c}
\ddot{x} \\
\ddot{y} \\
\ddot{z}
\end{array}\right]=\left[\begin{array}{c}
2 \omega \dot{z}+\dot{\omega} z+\omega^{2} x-\frac{\mu x}{|\boldsymbol{R}+\boldsymbol{r}|^{3}} \\
-\frac{\mu y}{|\boldsymbol{R}+\boldsymbol{r}|^{3}}\\
\omega^{2} z-2 \omega \dot{x}-\dot{\omega} x-\mu\left(\frac{z-R}{|\boldsymbol{R}+\boldsymbol{r}|^{3}}+\frac{1}{{R}^{2}}\right)
\end{array}\right]+\boldsymbol{u},
\label{eqn:nonlinear_space_rendezvous}
\end{align}
where $R:=|\boldsymbol{R}|,r:=|\boldsymbol{r}|$, $|\boldsymbol{R}+\boldsymbol{r}|^{2}:=x^{2}+y^{2}+(z-R)^{2}$, and $\omega$
is the orbital rate of the rotating coordinate system. Let $h$ be the orbital angular momentum of the target. Then, $R^{2} \omega=h=$ constant. Let $e \in[0,1)$ be the eccentricity of the target orbit, $\nu$ the true anomaly, $\rho=1+e \cos\nu,$ and
\begin{align}
k = \mu/h^{\frac{3}{2}}=\text{constant}.
\label{eqn:k}
\end{align}
The orbital rate $\omega$ satisfies
\begin{align}
\omega = h/ R^{2} =k^{2} \rho^{2}.
\end{align}
The eccentric anomaly $E$ and the true anomaly $\nu$ satisfy the following equations:
\begin{align}
    \sin (E)=\frac{\sqrt{1-e^{2}} \sin (\nu)}{1+e \cos (\nu)}, \;\; \cos (E)=\frac{e+\cos (\nu)}{1+e \cos (\nu)}
\end{align}
In addition, the eccentric anomaly $E$ and time $t$ satisfy the following well-known Kepler's equation:
\begin{align}
    t=\frac{T_o}{2 \pi}\big( E-e \sin (E) \big),
\end{align}
where $T_o$ is the time period of the orbit. The nonlinear equation given in \eqref{eqn:nonlinear_space_rendezvous} can be rewritten in state space form as follows:
\begin{align}
    \boldsymbol{\dot{x}}_c=\boldsymbol{f}(\boldsymbol{x}_c,\boldsymbol{u}),
    \label{eqn:continous_compact_spacecraft}
\end{align}
where $\boldsymbol{x}_c=[x\;y\;z\;\dot{x}\;\dot{y}\;\dot{z}]^\mathrm{T}$. The vectors $[x\;y\;z]^\mathrm{T}$ and $[\dot{x}\;\dot{y}\;\dot{z}]^\mathrm{T}$ correspond to, respectively, the position and velocity of the active chaser spacecraft with respect to the target spacecraft in the LVLH frame.

\subsection{Discrete-time nonlinear model}
A classical fourth order Runga Kutta discretization method \cite{butcher2008numerical_runga} is used to convert the continuous-time nonlinear dynamical system given by Eq. \eqref{eqn:continous_compact_spacecraft} to a discrete-time nonlinear dynamical system as follows:
\begin{align}
\boldsymbol{x}(k+1)&=\boldsymbol{x}(k)+\frac{T}{6}\left(\boldsymbol{k}_{\mid 1}+2 \boldsymbol{k}_{\mid 2}+2 \boldsymbol{k}_{\mid 3}+\boldsymbol{k}_{\mid 4}\right),
\label{eqn:discrete_runga}
\end{align}
where $k\in[0, N-1]_d$, $t_k=\frac{t_f}{N}k=Tk$, $t_f$ is the final time, $T>0$ is the sampling period, $\boldsymbol{k}_{\mid 1}$, $\boldsymbol{k}_{\mid 2}$, $\boldsymbol{k}_{\mid 3}$, and $\boldsymbol{k}_{\mid 4}$ are given as follows \cite{butcher2008numerical_runga}: 
\begin{subequations}
\begin{align}
& \boldsymbol{k}_{\mid 1}=\boldsymbol{f}(\boldsymbol{x}(k),\boldsymbol{u}(k)) \\
& \boldsymbol{k}_{\mid 2}=\boldsymbol{f}\left(\boldsymbol{x}(k)+\frac{T}{2} \boldsymbol{k}_{\mid 1},\boldsymbol{u}{(k)}\right) \\
& \boldsymbol{k}_{\mid 3}=\boldsymbol{f}\left(\boldsymbol{x}(k)+\frac{T}{2} \boldsymbol{k}_{\mid 2},\boldsymbol{u}{(k)}\right) 
\\
& \boldsymbol{k}_{\mid 4}=\boldsymbol{f}\left(\boldsymbol{x}(k)+T \boldsymbol{k}_{\mid 3},\boldsymbol{u}{(k)}\right), 
\end{align}
\end{subequations}
The state of the continuous-time system $\boldsymbol{x}_c$ and the state $\boldsymbol{x}$ of the discrete-time system are related as follows: $\boldsymbol{x}_c(t_k) \approx \boldsymbol{x}(k)$. From Eq. \eqref{eqn:discrete_runga}, the discrete nonlinear spacecraft rendezvous can be written in compact form as follows
\begin{align}
    \boldsymbol{x}(k+1)=\boldsymbol{h}(\boldsymbol{x}(k),\boldsymbol{u}(k)).
    \label{eqn:discrete_nonlinear_equation}
\end{align}
Now we present the problem, we address in this paper.

\begin{problem}
Given the discrete-time nonlinear rendezvous dynamics \eqref{eqn:discrete_nonlinear_equation}, $N>0$, the initial state $\boldsymbol{x}_0$ and the final $\boldsymbol{x}_f$, find the control input $\boldsymbol{u}^\star(k)$ for all $k\in[0,N-1]_d$ which will steer the active spacecraft from initial state $\boldsymbol{x}_0$ to final state $\boldsymbol{x}_f$ at $k=N$ while minimizing the following performance index:
\begin{align}
 J_{2,1}\left(\boldsymbol{u}^{\star}(k) \right):=\sum_{i=0}^{N-1}\|\boldsymbol{u}(i)\|_2. \label{eqn:performance_index_nonlinear}
\end{align}
\label{problem:koopman_operator}
\end{problem}

The solution to Problem \ref{problem:koopman_operator} poses significant challenges and requires the use of computationally expensive and sophisticated optimization algorithms \cite{arzelier2013new_conv4,ping_lu2013autonomous_cone_2,ping_liu2014solving,serra2018fuel,arzelier2016linearized_compare}. 
Instead of using these optimization algorithms, we propose the following two step approach for Problem \ref{problem:koopman_operator}. First, we use a Koopman based approach to approximate the discrete nonlinear model \eqref{eqn:discrete_nonlinear_equation} to a higher dimensional (lifted) linear state space model. Second, we exploit the linearity of this lifted state space model to solve the minimum-fuel problem for a thrust vectoring spacecraft.

\section{Koopman operator}\label{sec:koopman_operator}

\subsection{Quick review of Koopman operator}
Koopman operator $\mathcal{K}$ is an infinite dimensional operator which operates on a collection of observable functions $\boldsymbol{g}=[g_1,\;\;g_2,\;\dots, g_{N_k}]^\mathrm{T}$ where $g_{i}:\mathbb{R}^n\rightarrow\mathbb{R}$. The evolution of these set of functions is linear. In other words, the Koopman operator $\mathcal{K}: \mathcal{F} \rightarrow \mathcal{F}$ is defined as follows: 
\begin{align}
 (\mathcal{K} \boldsymbol{g})\boldsymbol{x}(k)=\boldsymbol{g}(\boldsymbol{f}(\boldsymbol{x}(k)))=\boldsymbol{g}(\boldsymbol{x}(k+1))  
\end{align}
 where $\mathcal{F}$ is a space of functions (often referred to as observables) which are invariant under the action of the Koopman operator. In contrast to the dynamics that are linearized around a fixed linearization point and become inaccurate away from this point, the Koopman operator describes the evolution of the observables of a nonlinear system with full accuracy throughout the state space.

The observable function $\boldsymbol{g}(\boldsymbol{x})$ can be written as 
\begin{align}
   \boldsymbol{g}(\boldsymbol{x})=[g_1(\boldsymbol{x}),\;g_2(\boldsymbol{x}),\dots {g}_{N_k}(\boldsymbol{x})]^\mathrm{T},
\end{align}
where $g_i(\boldsymbol{x}):\mathbb{R}^n\rightarrow\mathbb{R},\;\forall\;i\in\{1,\dots, N_k\}$, $N_k\gg n$. The state $\boldsymbol{z}$ is often referred to as the lifted state as it corresponds to the state of the system in the lifted state space which can be written in compact form as follows:
\begin{align}
    \boldsymbol{z}(k)=\boldsymbol{g}(\boldsymbol{x}(k)),
\end{align}
\subsection{Lifted dynamics for rendezvous operations}
In this section, we present the main steps for the approximation of the discrete-time nonlinear rendezvous equation \eqref{eqn:discrete_nonlinear_equation} with higher dimensional linear state space model using Koopman operator. Consider the discrete-time nonlinear rendezvous equation given in Eq. \eqref{eqn:discrete_nonlinear_equation}. Our goal is to approximate Eq. \eqref{eqn:discrete_nonlinear_equation} as the following linear lifted state space model
\begin{align}
    \boldsymbol{z}(k+1)=A_{\text{koop}}\boldsymbol{z}(k)+B_{\text{koop}}\boldsymbol{u}(k),
    \label{eqn:koopman_linear_dynamics}
\end{align}
 where $N_k$ is the dimension of the lifted state $\boldsymbol{z}(k)$, $A_{\text{koop}} \in \mathbb{R}^{N_k \times N_k}$, $B_{\text{koop}} \in \mathbb{R}^{N_k \times m}$,  $\boldsymbol{z}(k)\in\mathbb{R}^{N_k}$, $\boldsymbol{u}(k)\in\mathbb{R}^m$ and $k\in[0,N-1]_d$.
The initial condition $\boldsymbol{z}_0$ is given by
\begin{align}
  \boldsymbol{z}_0= \boldsymbol{g}(\boldsymbol{x}_0)=[g_1(\boldsymbol{x}_0),\;\;\;g_2(\boldsymbol{x}_0),\dots, {g}_{N_k}(\boldsymbol{x}_0)]^\mathrm{T}
  \label{eqn:koopman_initial_condition}
\end{align}
where $\boldsymbol{x}_0=\boldsymbol{x}(0)$ is the initial condition for the original discrete nonlinear equation given in Eq. \eqref{eqn:discrete_nonlinear_equation}. 
The terminal state of the lifted space dynamics given in Eq.~\eqref{eqn:koopman_linear_dynamics} can be written as 
\begin{align}
    \boldsymbol{z}(N)=A_{\text{koop}}^{N} \boldsymbol{z}_{0}+\sum_{\tau=0}^{N-1} A_{\text{koop}}^{N-1-\tau} B_{\text{koop}} \boldsymbol{u}(\tau),
    \label{eqn:terminal_state}
\end{align}
where $\boldsymbol{z}(N)= \boldsymbol{g}(\boldsymbol{x}(N))$ and $\boldsymbol{x}(N)$ denotes the terminal state of the original discrete-time nonlinear equation \eqref{eqn:discrete_nonlinear_equation}. The terminal state can be rewritten in a compact form as follows
\begin{align}
    \boldsymbol{z}(N)=\boldsymbol{C}_{{N}_{\text{koop}}} \boldsymbol{u}_{\text{koop}}+\boldsymbol{\beta}_{\text{koop}},
    \label{eqn:terminal_state_koopman}
\end{align}
where $\boldsymbol{C}_{{N}_{\text{koop}}}\in\mathbb{R}^{{N}_{\text{k}}\times Nm}$, $\boldsymbol{u}_{\text{koop}}\in\mathbb{R}^{Nm}$ and $\boldsymbol{\beta}_{\text{koop}}\in\mathbb{R}^{{N}_{\text{k}}}$ are defined as
\begin{subequations}
\begin{align}
    & \boldsymbol{u}_{\text{koop}} :=[\boldsymbol{u}(0)^\mathrm{T},\;\boldsymbol{u}(1)^\mathrm{T},\;\dots,\boldsymbol{u}(N-1)^\mathrm{T}]^\mathrm{T},\\
    & \boldsymbol{C}_{{N}_{\text{koop}}} :=[A_{\text{koop}}^{N-1} {B_{\text{koop}}}, \ldots, {B_{\text{koop}}}], \label{eqn:c_n_koopamn}\\ 
    & \boldsymbol{\beta}_{\text{koop}} :=A_{\text{koop}}^{N} \boldsymbol{z}_{0}. \label{eqn:beta_koopman}
\end{align}
\end{subequations}

\subsection{A data-driven method to compute $A_\text{koop}$ and $B_\text{koop}$}\label{sec:data_driven}
The given discrete-time nonlinear spacecraft rendezvous dynamics is entirely known from Eq. \eqref{eqn:discrete_nonlinear_equation}. We now use a data-driven approach to approximate the matrices $A_\text{koop}$ and $B_\text{koop}$ that appear in \eqref{eqn:koopman_linear_dynamics}. To this aim, a set of random control inputs and random set of initial states $\boldsymbol{x}_0$ are chosen whose entries are drawn from a uniform distribution $[-1,1]$. These randomly generated control inputs are applied sequentially to Eq. \eqref{eqn:discrete_nonlinear_equation} with initial state $\boldsymbol{x}_0$ to get the subsequent states. Let the control input $\boldsymbol{u}(k)$ be applied to take the state of the active spacecraft from $\boldsymbol{x}(k)$ to $\boldsymbol{x}(k+1)$. In this way, we construct the matrices $\boldsymbol{X},\boldsymbol{U},$ and $\boldsymbol{Y}$ where states $\boldsymbol{X}=[\boldsymbol{x}(0),\dots,\boldsymbol{x}(d)]$ along with their respective control inputs are stored $\boldsymbol{U}=[\boldsymbol{u}(0),\dots,\boldsymbol{u}(d)]$ and let $\boldsymbol{Y}=[\boldsymbol{x}(1),\dots,\boldsymbol{x}({d+1})]$ where $(d+1)$ is the number of data points. 
The matrix $\boldsymbol{Y}$ can be expressed as follows:
\begin{align}
  \boldsymbol{Y}=\boldsymbol{f}(\boldsymbol{X},\boldsymbol{U}). 
\end{align}

Given the data $\boldsymbol{X}, \boldsymbol{Y},$ and $ \boldsymbol{U}$, the matrices $A_\text{koop}$ and $B_\text{koop}$ in \eqref{eqn:koopman_linear_dynamics} are obtained via the solution to the following least squares optimization problem:
\begin{align}
 \min _{A_\text{koop}, B_\text{koop}}\left\|\boldsymbol{Y}_{\mathrm{lift}}-A_\text{koop} \boldsymbol{X}_{\mathrm{lift}}-B_\text{koop} \boldsymbol{U}\right\|_{F}  ,
 \label{eqn:least_squares_optimization}
\end{align}
where 
\begin{align}
  &\boldsymbol{X}_{\mathrm{lift}}=\left[\boldsymbol{g}(\boldsymbol{x}({0})), \ldots, \boldsymbol{g}(\boldsymbol{x}({d}))\right],\\ &\boldsymbol{Y}_{\mathrm{lift}}=\left[\boldsymbol{g}(\boldsymbol{x}({1})), \ldots, \boldsymbol{g}(\boldsymbol{x}({d+1}))\right], 
\end{align}
with
\begin{align}
 \boldsymbol{g}(\boldsymbol{x})=[{g}_{1}(\boldsymbol{x}), \dots, {g}_{{N}_k}(\boldsymbol{x})]^\mathrm{T} ,  
\end{align}
being a given collection of nonlinear observable functions $g_i(\boldsymbol{x})$ for all $i\in\{1,\dots,N_k\}$. The symbol $\|\cdot\|_{F}$ denotes the Frobenius norm of a matrix. 
The analytical solution to \eqref{eqn:least_squares_optimization} is given by:
\begin{align}
[A_\text{koop}, B_\text{koop}]=\boldsymbol{Y}_{\text {lift }}\left[\boldsymbol{X}_{\text {lift}}, \boldsymbol{U}\right]^{\dagger}.   
\end{align}
where $(.)^{\dagger}$ denotes the Moore-Penrose pseudoinverse operator.

\section{Proposed solution approach for the minimum fuel problem based on the IRLS algorithm}\label{sec:algorithms}
Now that we have approximated the matrices $A_\text{koop}$ and $B_\text{koop}$ of the lifted space dynamics \eqref{eqn:koopman_linear_dynamics}, a modified version of Problem \ref{problem:koopman_operator} is presented next. 
\begin{problem}
Let $\boldsymbol{x}_{0}$, $\boldsymbol{x}_f\in\mathbb{R}^6$ and $N>0$ be given. Find a control sequence $\boldsymbol{u}_{\text{koop}}^\star(k)\in\mathbb{R}^3$ for all $k\in[0,N-1]_d$ that will minimize the performance index given in \eqref{eqn:performance_index_nonlinear} and 
subject to the following terminal equality constraint.
\begin{align}
    \boldsymbol{C}_{{N}_{\text{koop}}} \boldsymbol{u}_{\text{koop}}+\boldsymbol{\beta}_{\text{koop}}=\boldsymbol{z}_f,
    \label{eqn:problem_2_terminal}
\end{align}
where $\boldsymbol{z}_f=\boldsymbol{g}(\boldsymbol{x}(N))$.
\label{problem:optimal_control_koopman}
 \end{problem} 
The proposed approach to solve Problem \ref{problem:optimal_control_koopman} is based on the an iterative approach known as the Iteratively Reweighted Least Squares (IRLS) algorithm. It is a popular tool for the computation of the minimum $\ell_2/\ell_1$ or $\ell_1$ norm solution to an under-determined linear system in the literature of compressive sensing \cite{b:Foucart2013}.
\subsection{IRLS Algorithm}
The iterative approach presented here computes an approximate solution to the minimum $\ell_2/\ell_1$ norm problem in closed form via the solution of a corresponding sequence of convex quadratic programs. In particular, at every iteration $j$, $\boldsymbol{u}_{\text{koop}}^{[j+1]}$ corresponds to the solution of the following convex quadratic program:
\begin{align*}
& \text{(QP):}~~~\underset{\boldsymbol{u}}{\min}\sum_{i=0}^{N-1} \sum_{k=1}^{m} \boldsymbol{u}(i)^{\mathrm{T}}\boldsymbol{w}^{[j]}(k)\boldsymbol{u}(i)~~ \text{subject to}\;\;\eqref{eqn:problem_2_terminal}
\end{align*}
with
\[
\boldsymbol{w}^{[j]}:=[w^{[j]}(0)^\mathrm{T},w^{[j]}(1)^\mathrm{T}, \dots, w^{[j]}(N-1)^\mathrm{T}]^\mathrm{T}\in\mathbb{R}_{>0}^{Nm}, 
\] 
where $w^{[j]}(k)\in\mathbb{R}^m$ for all $k\in[0,N-1]_d$ and $m=3$ for a thrust vectoring spacecraft~\cite{leomanni2019sum}. First, the input parameters $\boldsymbol{w}^{[0]}(k)$ for all $k\in[1,Nm]$ and  $\varepsilon^{[0]}$ are initialized to 1 and $j$ is set to zero. We define the weight matrix 
\begin{align}
  \hspace{-0.3cm}\mathbf{W}^{[j]}(k)=\operatorname{diag}\left(\boldsymbol{w}^{[j]}{(k m+1)}, \dots,\boldsymbol{w}^{[j]}{(k m+m)}\right) ,
  \label{eqn:weight_matrix_l2}
\end{align}
for $k\in[0,N-1]_d$, which is a positive definite matrix provided that $\boldsymbol{w}^{[j]}\geq0$. Furthermore, let  
\begin{align}
  \boldsymbol{\mathcal{W}}^{[j]}=\operatorname{bdiag}\left(\mathbf{W}^{[j]}(0), \ldots, \mathbf{W}^{[j]}(N-1)\right). 
  \label{eqn:weight_matirx_big_l2}
\end{align}
%
\begin{figure*}[ht]
\captionsetup[subfigure]{justification=centering}
\centering
\begin{subfigure}{0.3\textwidth}
{\includegraphics[scale=0.28]{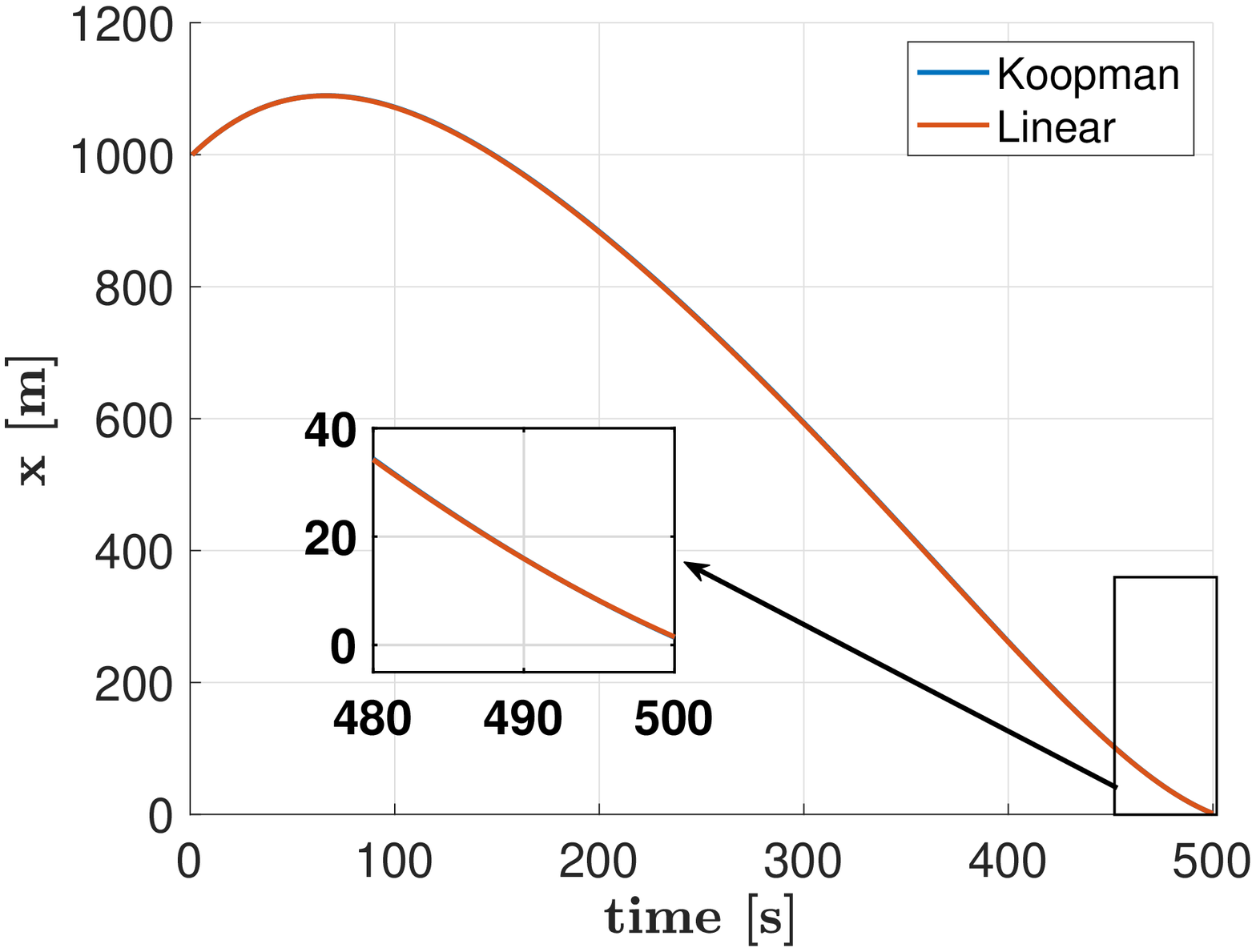}}
\caption{$x$}
\label{fig:}
\end{subfigure}
\begin{subfigure}{0.3\textwidth}
\includegraphics[scale=0.28]{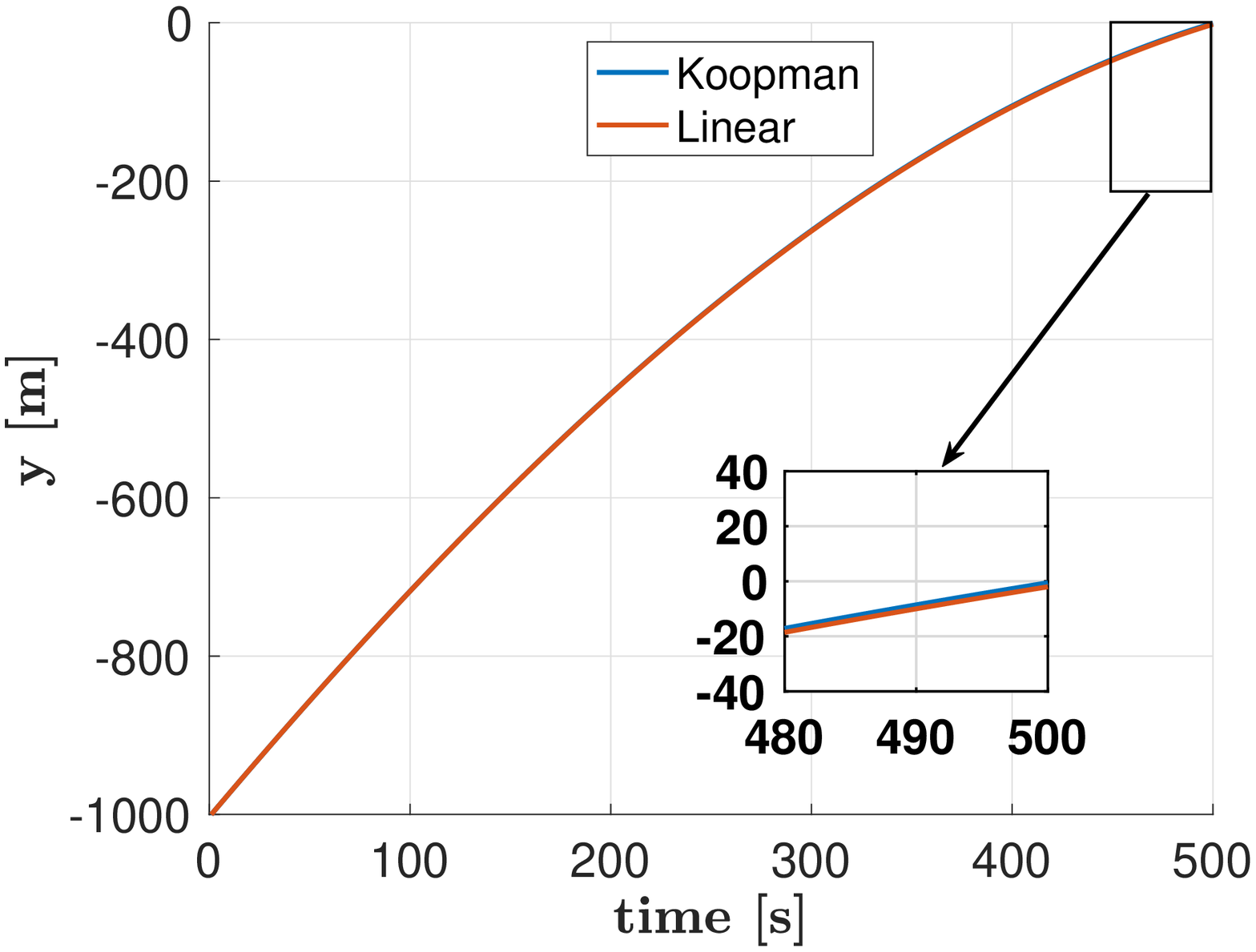}
\caption{$y$}
\label{fig:}
\end{subfigure}
\begin{subfigure}{0.3\textwidth}
\includegraphics[scale=0.28]{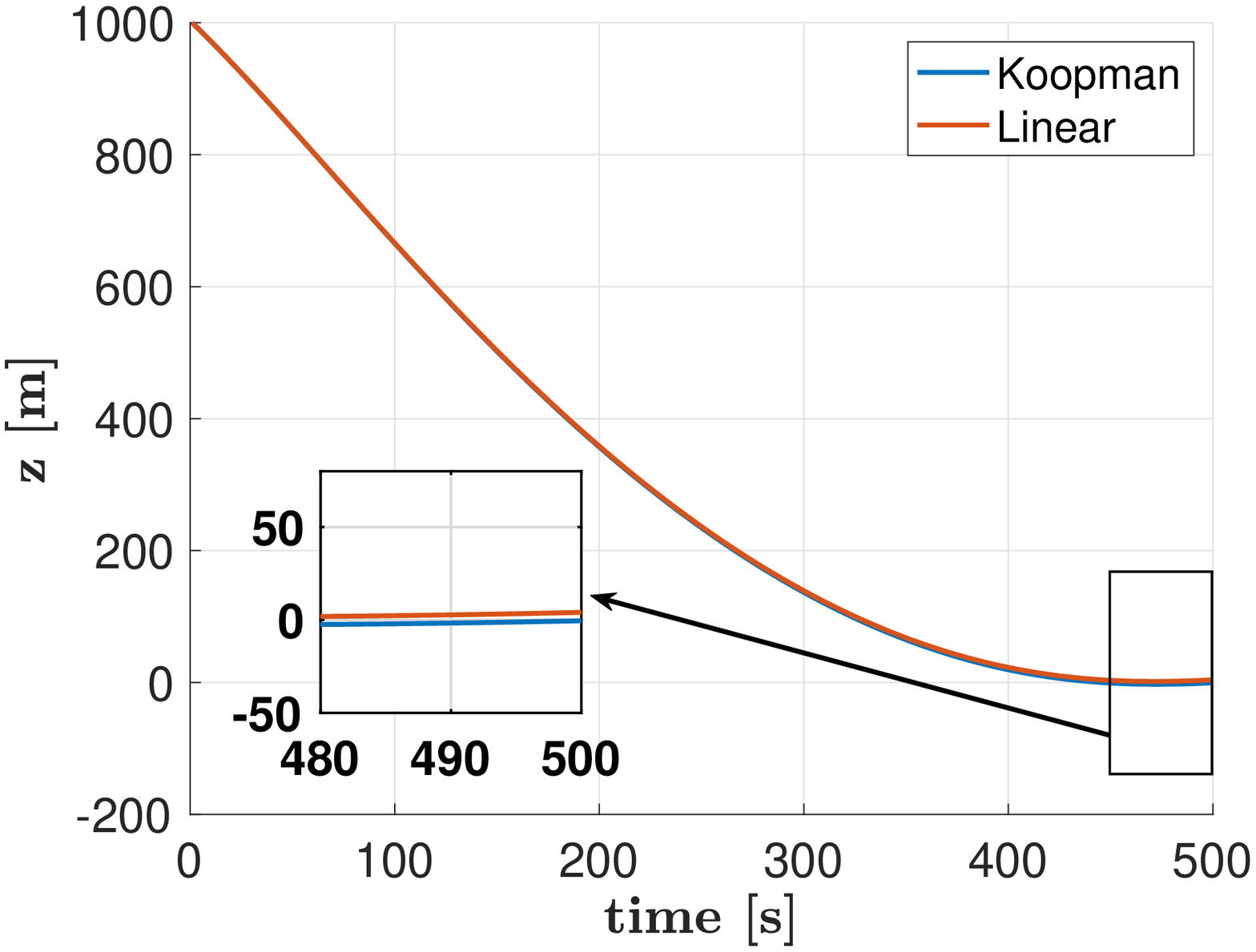}
\caption{$z$}
\label{fig:}
\end{subfigure}
\begin{subfigure}{0.3\textwidth}
{\includegraphics[scale=0.28]{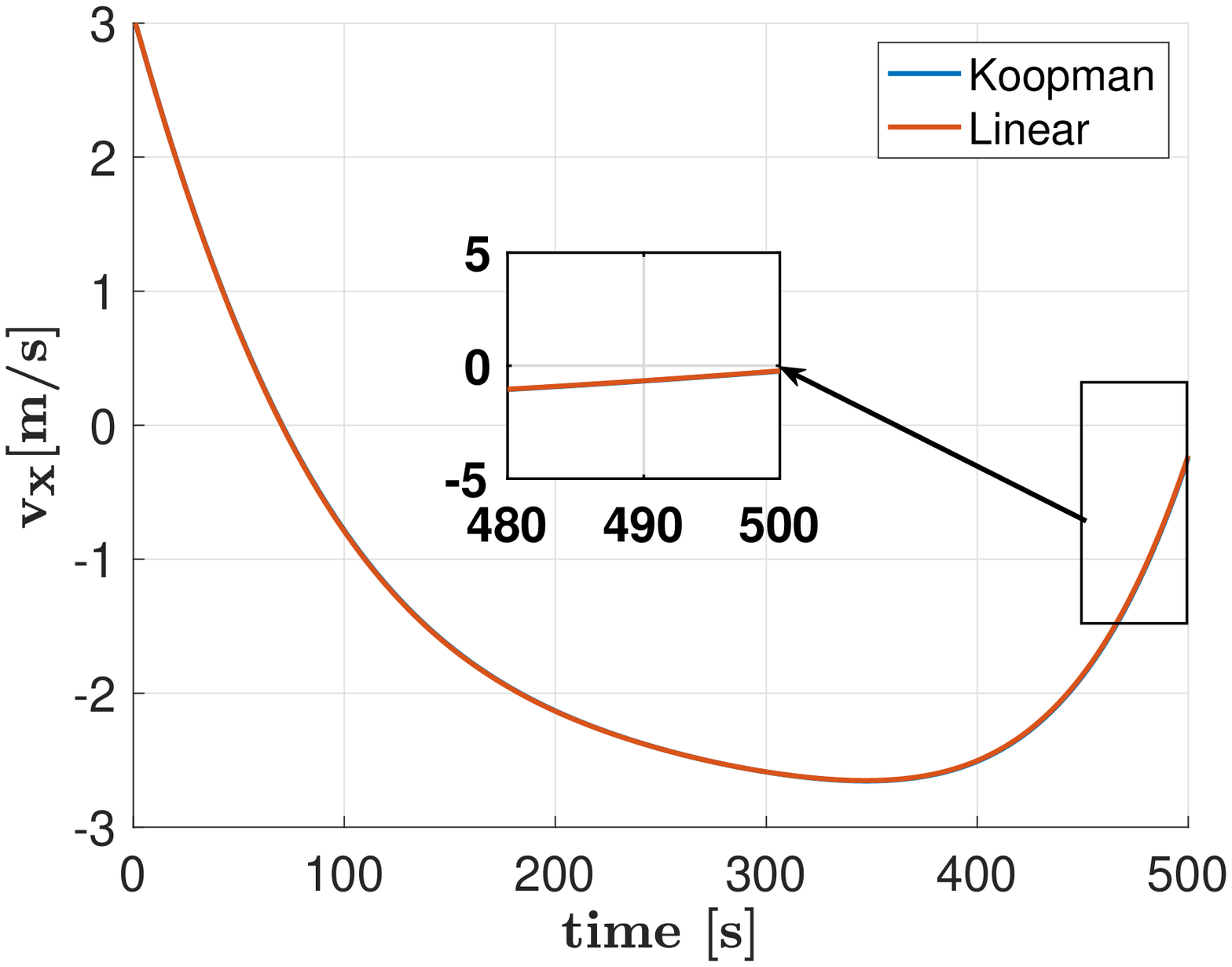}}
\caption{$\dot{x}$}
\label{fig:}
\end{subfigure}
\begin{subfigure}{0.3\textwidth}
\includegraphics[scale=0.28]{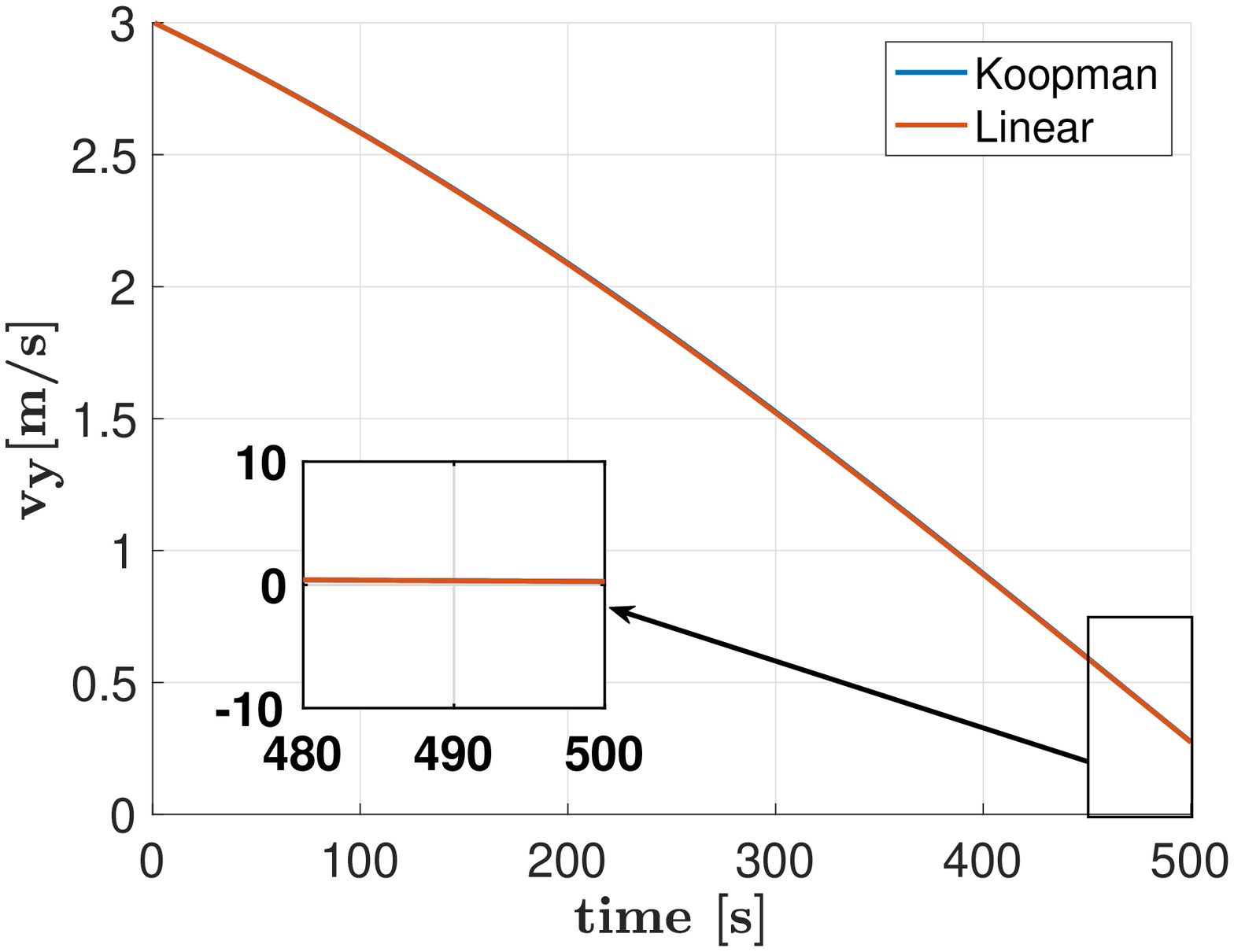}
\caption{$\dot{y}$}
\label{fig:}
\end{subfigure}
\begin{subfigure}{0.3\textwidth}
\includegraphics[scale=0.28]{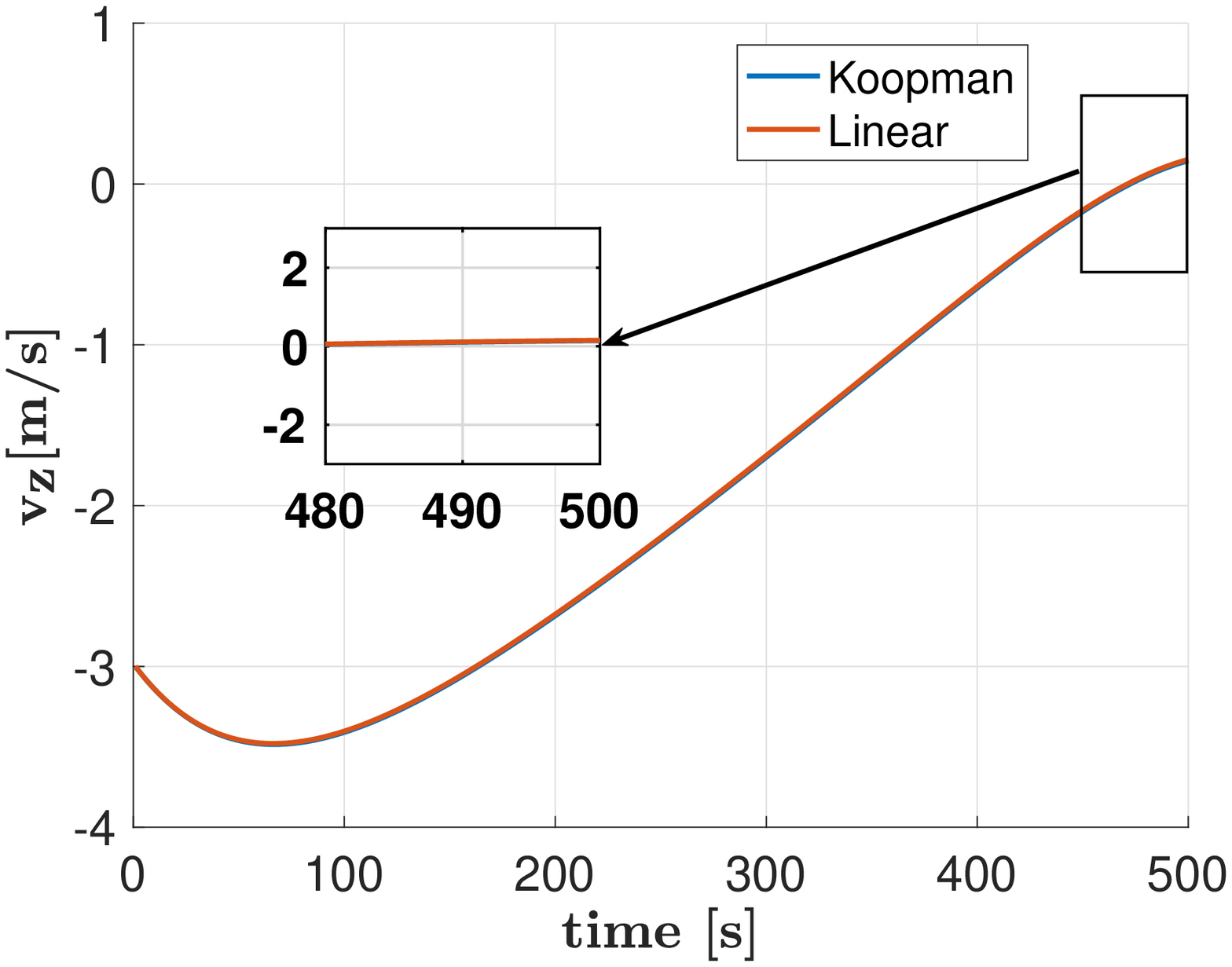}
\caption{$\dot{z}$}
\label{fig:}
\end{subfigure}
\caption{Evolution of states for short field rendezvous. In this case, the control inputs $\boldsymbol{u}_{\text{lin}}$ and $\boldsymbol{u}_\text{koop}$ generated using the linearized dynamics \eqref{eqn:terminal_constraint_linear_cn_beta} and the lifted space dynamics \eqref{eqn:terminal_state_koopman} respectively are able to steer the active spacecraft from initial state $\boldsymbol{x}_0$ to final state $\boldsymbol{x}_f$ with comparable accuracy. However, as seen from Table \ref{tab:terminal_state_error}, the Koopman operator based approach gives better performance in terms of the terminal state error.}
\label{fig:states_evolution_short_rendezvous}
\end{figure*}
\begin{figure*}[]
\captionsetup[subfigure]{justification=centering}
\centering
\begin{subfigure}{0.41\textwidth}
\includegraphics[scale=0.35]{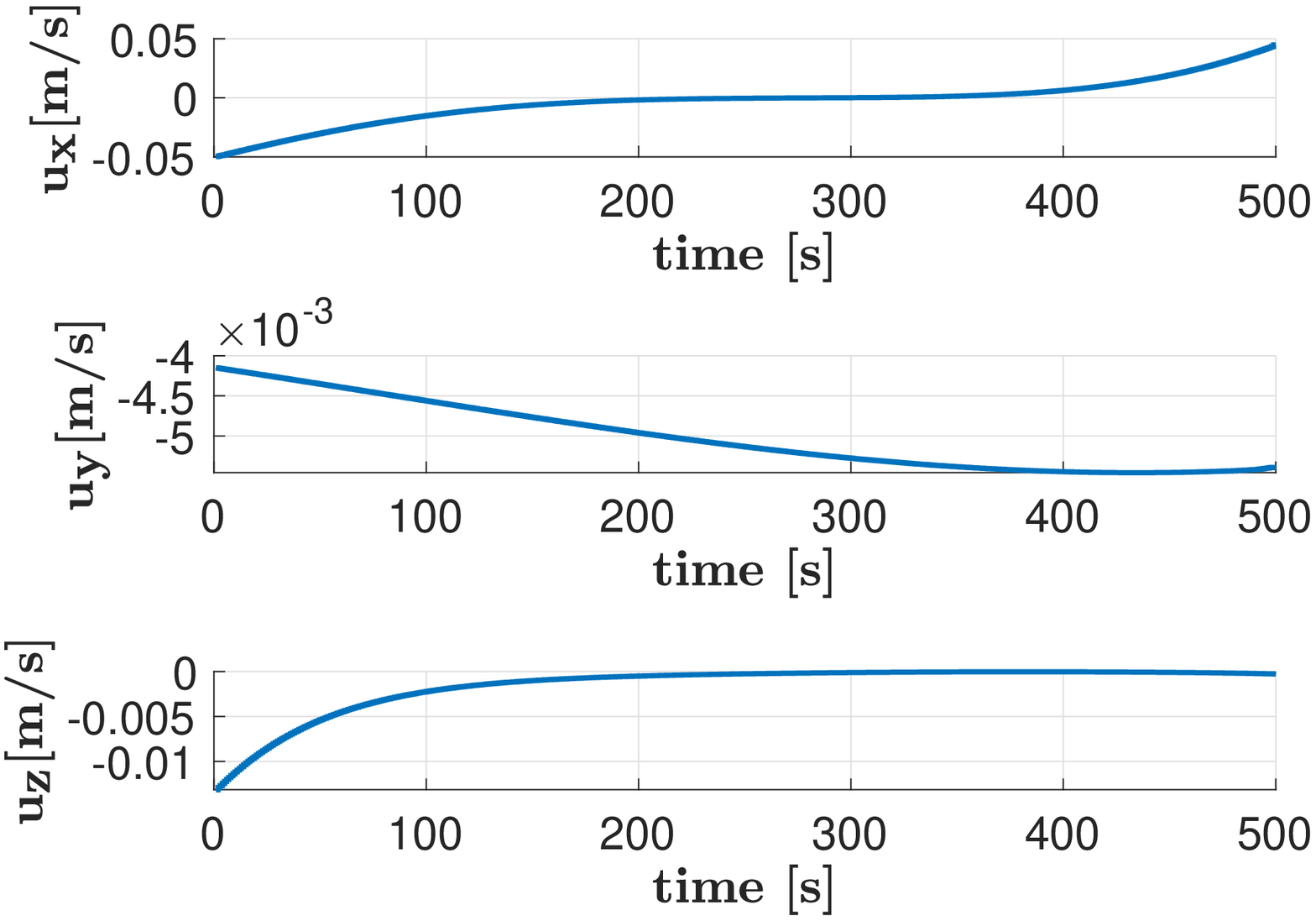}
\caption{$\boldsymbol{u}_\text{koop}$ for short-field rendezvous}
\label{fig:}
\end{subfigure}
\begin{subfigure}{0.41\textwidth}
\includegraphics[scale=0.35]{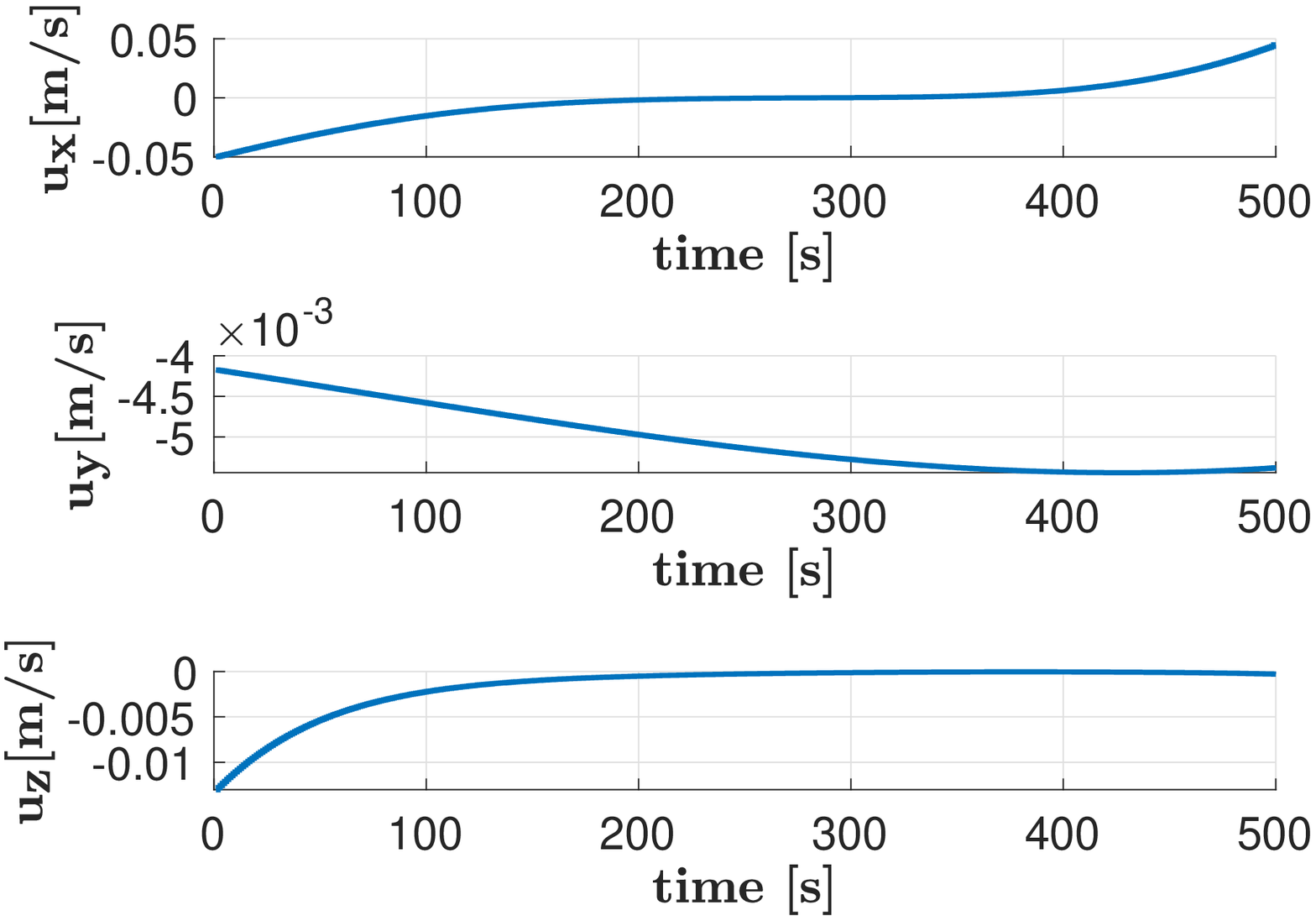}
\caption{$\boldsymbol{u}_{\text{lin}}$ for short-field rendezvous}
\label{fig:}
\end{subfigure}
\caption{Control inputs for short-field rendezvous}
\label{fig:}
\end{figure*}
\begin{algorithm}[H]
\caption{ IRLS algorithm for solving $\ell_2/\ell_1$ optimization problem}
\begin{algorithmic}[1]
\State $\boldsymbol{w}^{[0]}(i)=1\;\forall \;i\in[1,Nm]_d$
\State $\epsilon^{[0]}=1$
\For{$j=0\;\text{to}\;j_\text{max}$}
\For{$k=0,\dots,N-1$}
\State $\hspace{-0.4cm}\mathbf{W}^{[j]}(k)=\operatorname{diag}\left(\boldsymbol{w}^{[j]}{(k m+1)} \dots\boldsymbol{w}^{[j]}{(k m+m)}\right)$
\EndFor
\State $\boldsymbol{\mathcal{W}}^{[j]}=\operatorname{bdiag}\left(\mathbf{W}^{[j]}(0), \ldots, \mathbf{W}^{[j]}(N-1)\right)$
\State $\boldsymbol{u}_{\text{koop}}^{[j+1]}=(\boldsymbol{\mathcal{W}}^{[j]})^{-1}(\boldsymbol{C}_{N_{\text{koop}}}^{\mathrm{T}}(\boldsymbol{\mathcal{W}}^{[j]})^{-1})^{\mathrm{T}}\newline
\hspace{0.3cm}\; (\boldsymbol{C}_{N_{\text{koop}}}^{\mathrm{T}}(\boldsymbol{\mathcal{W}}^{[j]})^{-1}+\mathbf{I})^{-1}(\boldsymbol{C}_{N_{\text{koop}}}^{\mathrm{T}}(\boldsymbol{\mathcal{W}}^{[j]})^{-1})^{\mathrm{T}}\boldsymbol{\beta}_{\text{koop}}$
\State $ \varepsilon^{[j+1]}=\min \left\{\varepsilon^{[j]}, \|\boldsymbol{u}_{\text{koop}}^{[j+1]}\|_\infty\right\}$
\For{$\ell=1,\dots,Nm$}
\State $\boldsymbol{w}^{[j+1]}{(\ell)}=\left(\left(\boldsymbol{u}_{{\text{koop}}}^{[j+1]}(\ell)\right)^{2}+\left(\varepsilon^{[j+1]}\right)^{2}\right)^{-1/4}$
\EndFor
\If{$\epsilon\in[0,\bar{\epsilon}]$}
\State $\text{report}\; \say{\text{success}}$
\EndIf
\EndFor
\If{$\epsilon\notin[0,\bar{\epsilon}]$}
\State $\text{report}\; \say{\text{failure}}$
\EndIf
\end{algorithmic}
\label{algo:irls_sparse_l2/l1}
\end{algorithm}
Then, the solution $\boldsymbol{u}_{\text{koop}}^{[j+1]}$ to the (QP) is given by
\begin{align}
\boldsymbol{u}^{[j+1]}_\text{koop}=&(\boldsymbol{\mathcal{W}}^{[j]})^{-1}(\boldsymbol{C}_{{N}_\text{koop}}^{\mathrm{T}}
(\boldsymbol{\mathcal{W}}^{[j]})^{-1})^{\mathrm{T}}\nonumber\\
(\boldsymbol{C}_{{N}_\text{koop}}^{\mathrm{T}}&(\boldsymbol{\mathcal{W}}^{[j]})^{-1}+ \mathbf{I})^{-1}
(\boldsymbol{C}_{{N}_\text{koop}}^{\mathrm{T}}(\boldsymbol{\mathcal{W}}^{[j]})^{-1})^{\mathrm{T}} \boldsymbol{\beta}_\text{koop},
  \label{eqn:control_vector_l2}
\end{align}
where $\boldsymbol{C}_{N_\text{koop}}$ and $\boldsymbol{\beta}_\text{koop}$ are given by Eqs.~\eqref{eqn:c_n_koopamn} and~\eqref{eqn:beta_koopman} respectively. 
The weight matrices $\mathbf{W}^{[j]}(k)$ and $\boldsymbol{\mathcal{W}}^{[j]}(k)$ are updated at every iteration and are used to compute to control sequence $\boldsymbol{u}_\text{koop}$ at every iteration. This control sequence ultimately converges to the optimal control sequence $\boldsymbol{u}^\star_\text{koop}$ after a certain number of iterations that minimizes the $\ell_2/\ell_1$ norm and solves Problem \eqref{problem:optimal_control_koopman}. 
\begin{figure*}[]
\captionsetup[subfigure]{justification=centering}
\centering
\begin{subfigure}{0.3\textwidth}
{\includegraphics[scale=0.28]{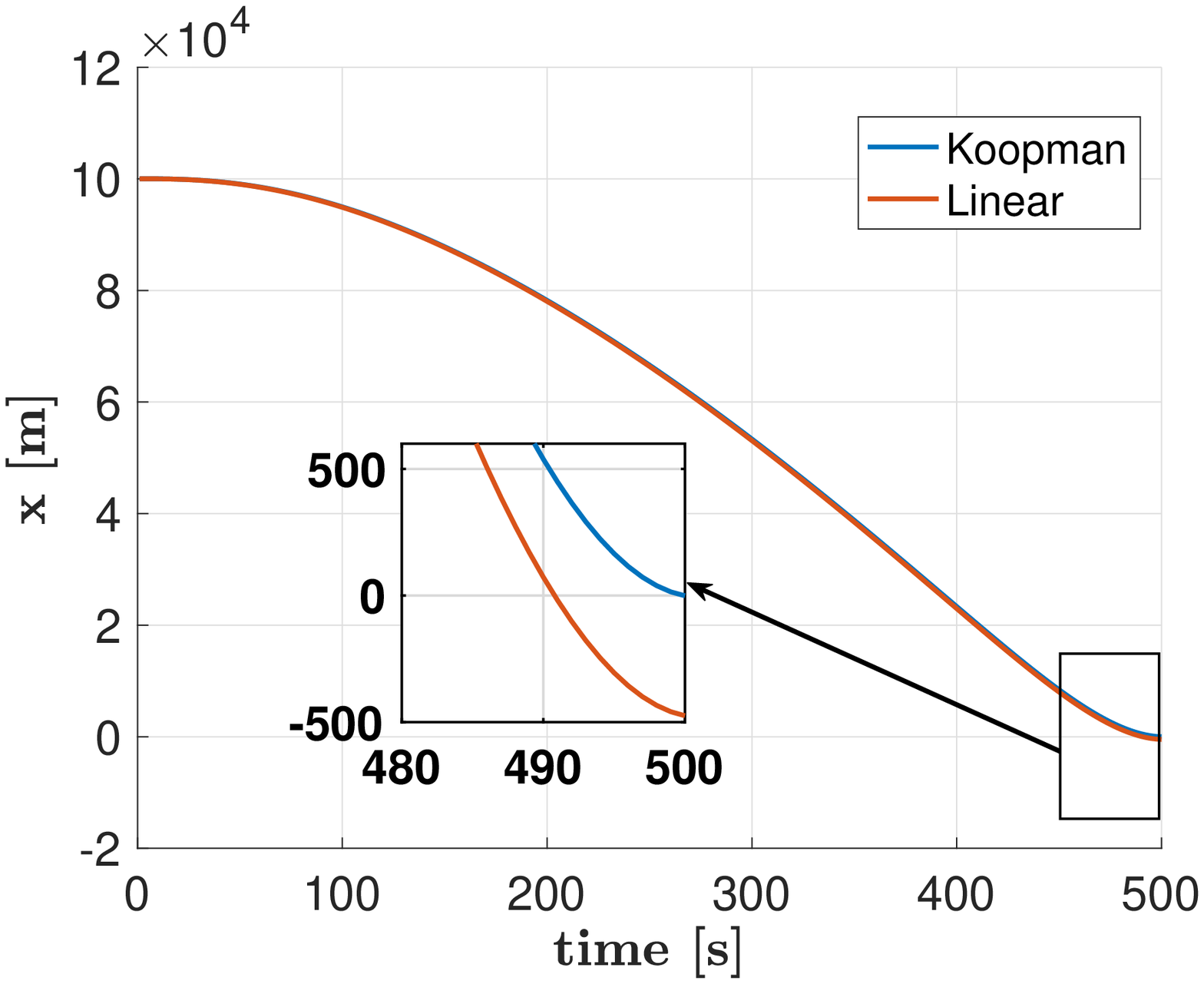}}
\caption{$x$}
\label{fig:}
\end{subfigure}
\begin{subfigure}{0.3\textwidth}
\includegraphics[scale=0.28]{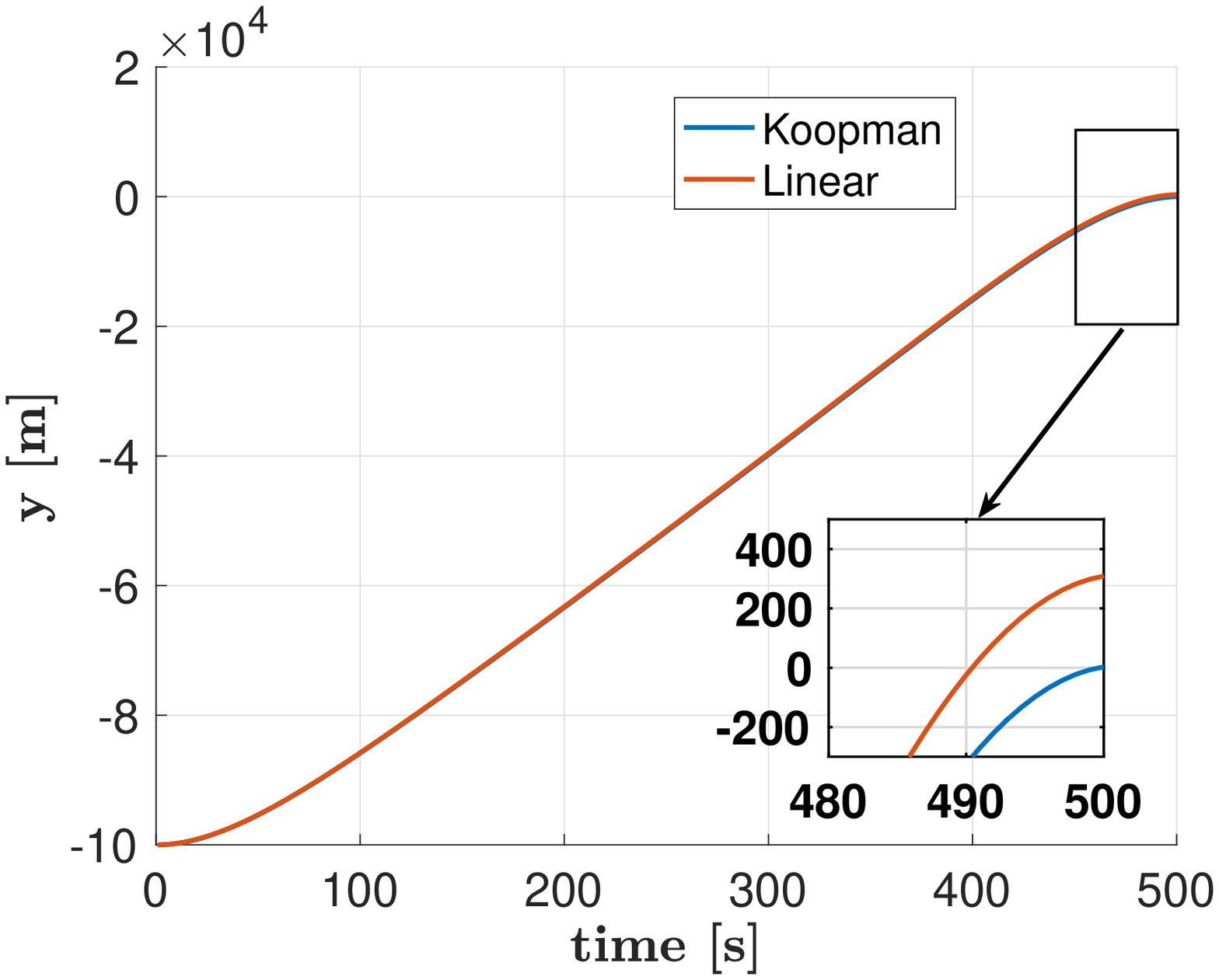}
\caption{$y$}
\label{fig:}
\end{subfigure}
\begin{subfigure}{0.3\textwidth}
\includegraphics[scale=0.28]{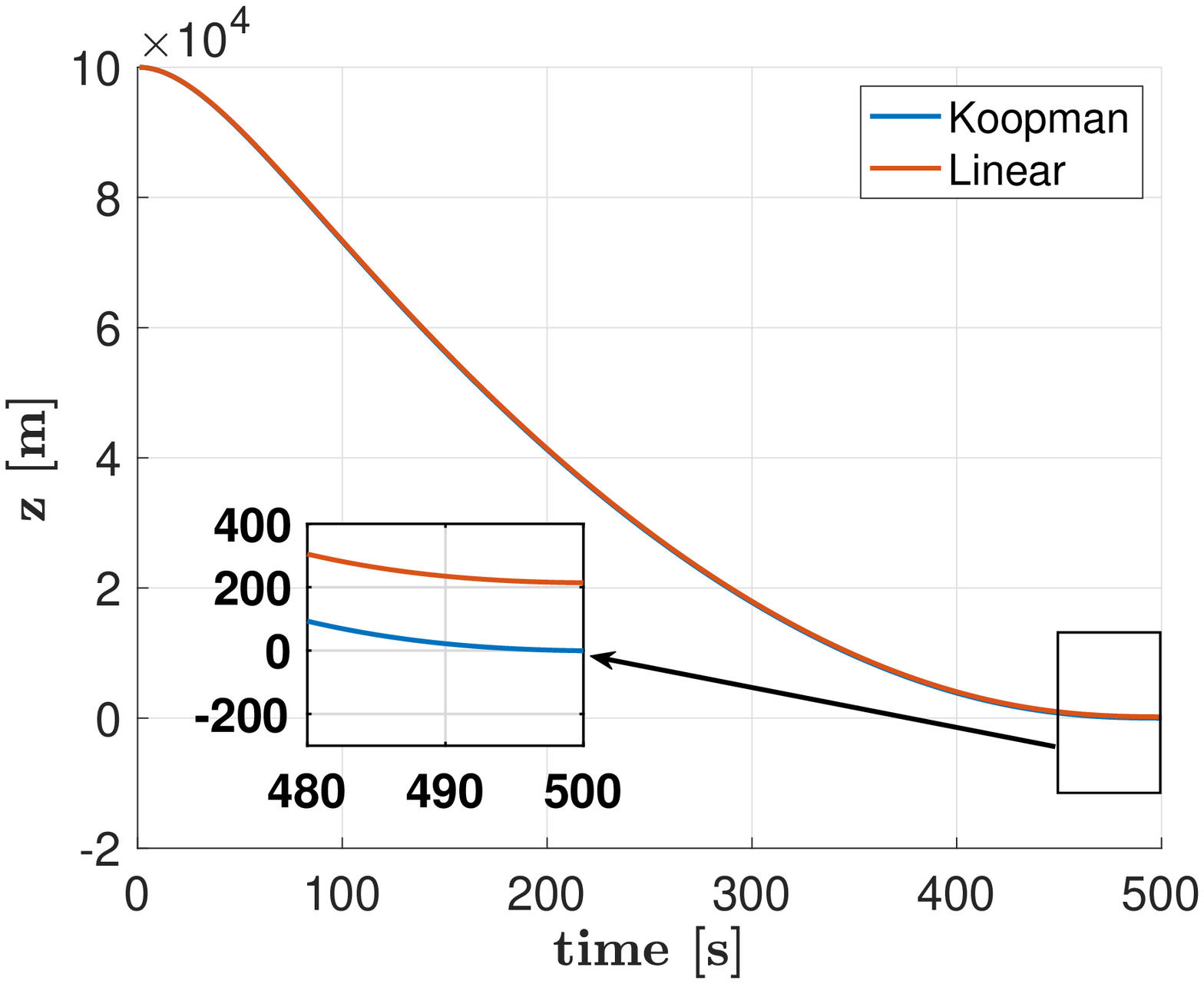}
\caption{$z$}
\label{fig:}
\end{subfigure}
\begin{subfigure}{0.3\textwidth}
{\includegraphics[scale=0.28]{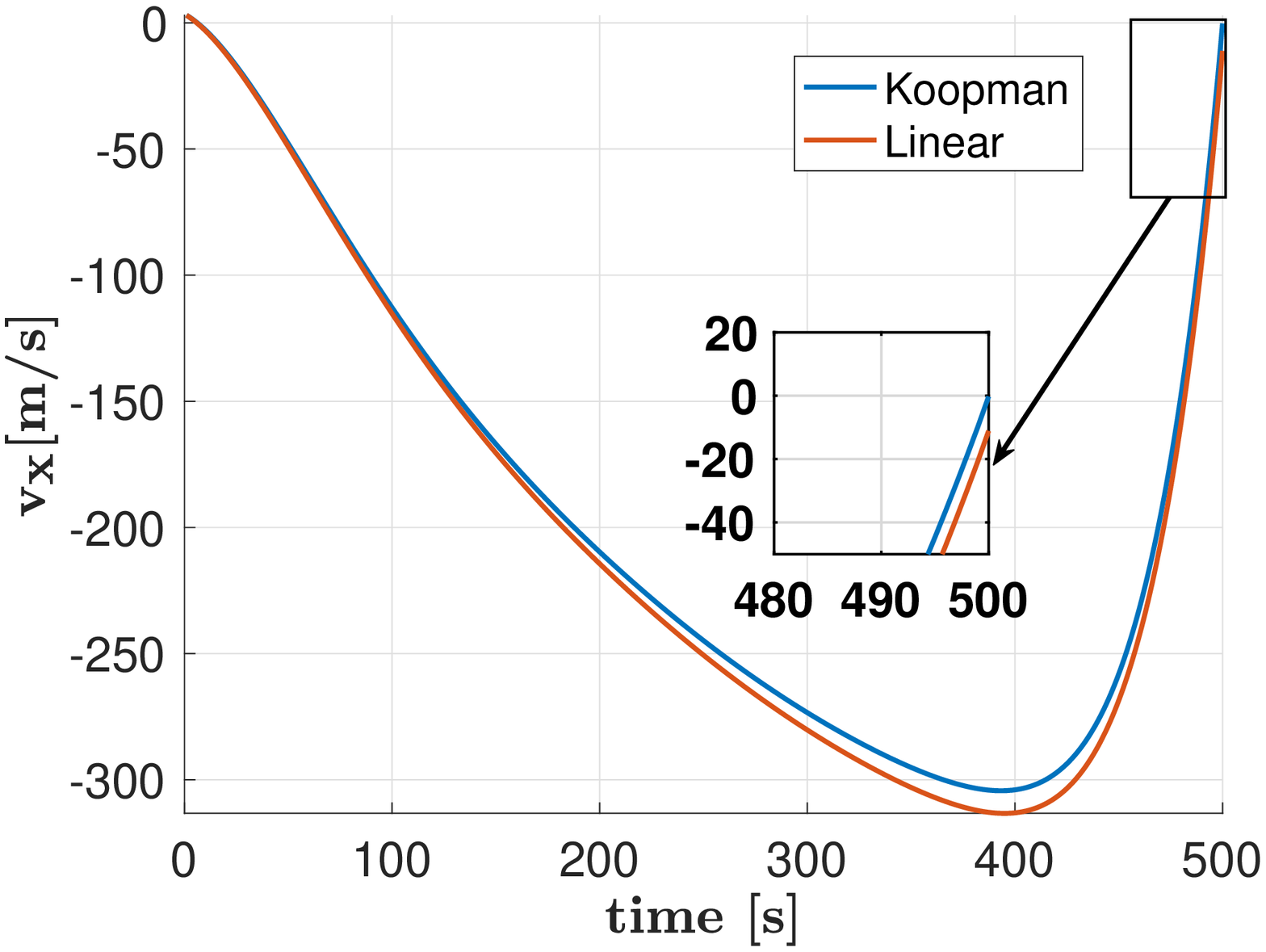}}
\caption{$\dot{x}$}
\label{fig:}
\end{subfigure}
\begin{subfigure}{0.3\textwidth}
\includegraphics[scale=0.28]{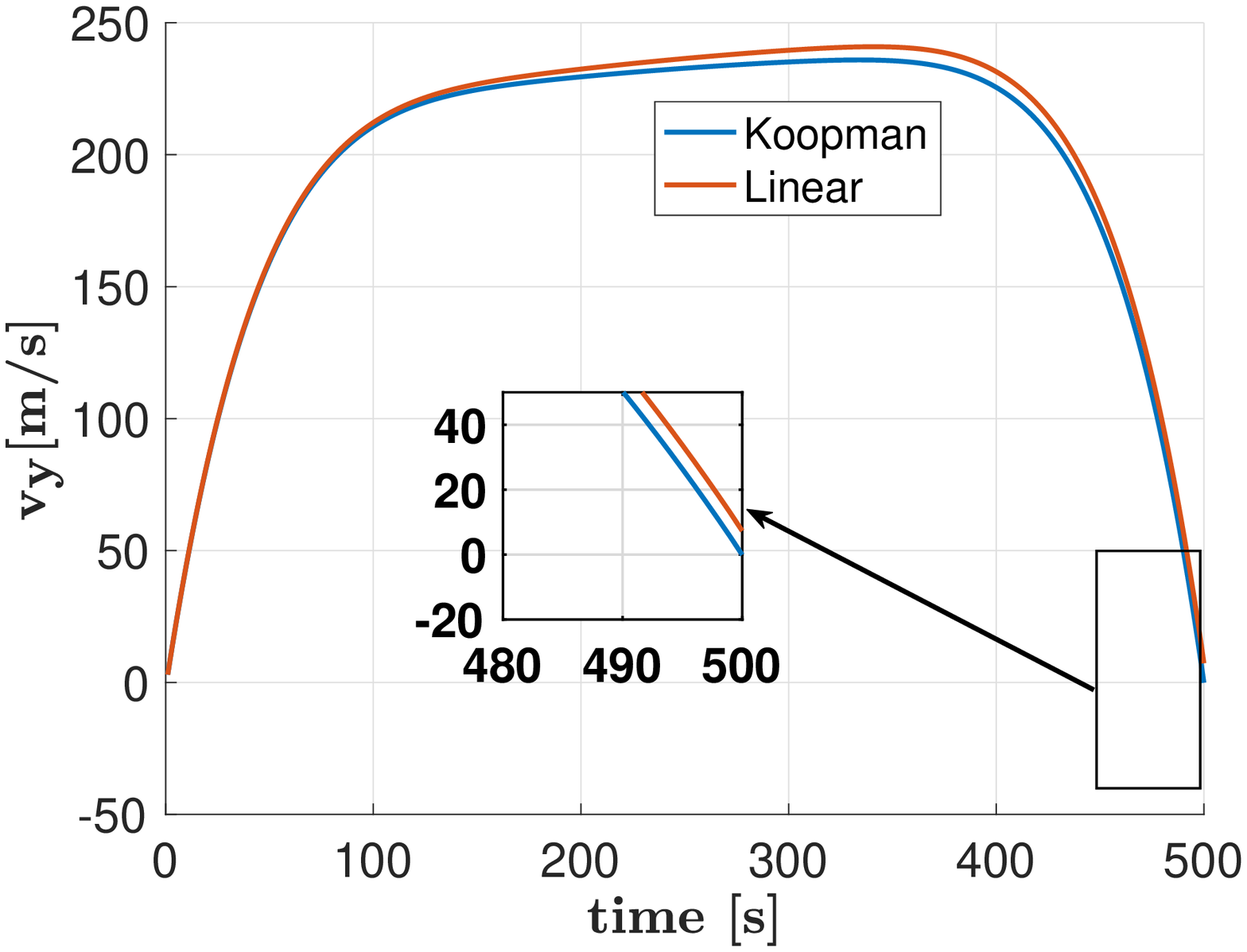}
\caption{$\dot{y}$}
\label{fig:}
\end{subfigure}
\begin{subfigure}{0.3\textwidth}
\includegraphics[scale=0.28]{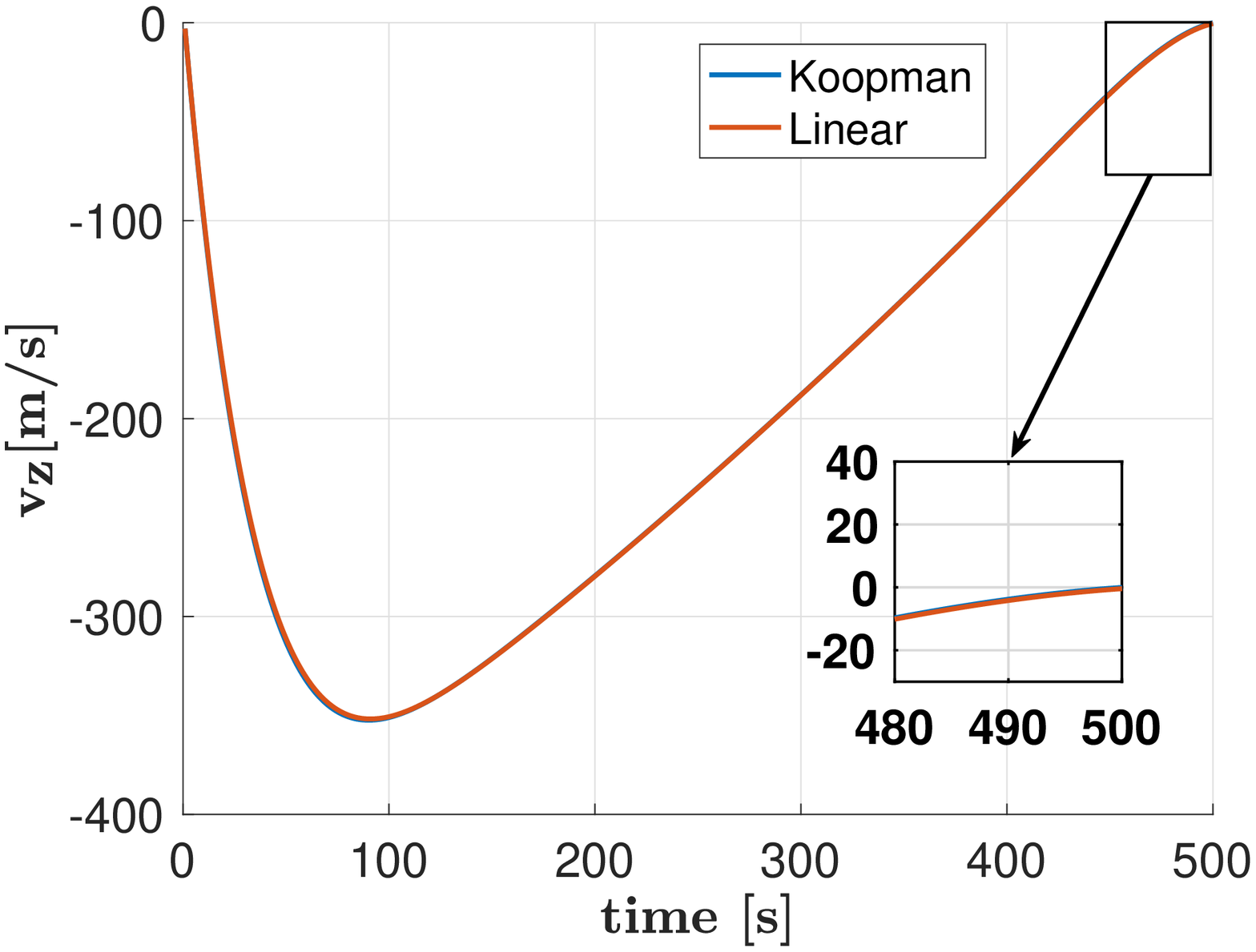}
\caption{$\dot{z}$}
\label{fig:}
\end{subfigure}
\caption{Evolution of states for far-field rendezvous. The control input $\boldsymbol{u}_\text{koop}$ generated using the lifted space dynamics \eqref{eqn:terminal_state_koopman} is able to steer the active spacecraft from initial state $\boldsymbol{x}_0$ to final state $\boldsymbol{x}_f$ with better accuracy than the control input $\boldsymbol{u}_{\text{lin}}$ which is generated using the linear dynamics.}
\label{fig:states_evolution_far_rendezvous}
\end{figure*}
\begin{figure*}[]
\captionsetup[subfigure]{justification=centering}
\centering
\begin{subfigure}{0.41\textwidth}
\includegraphics[scale=0.35]{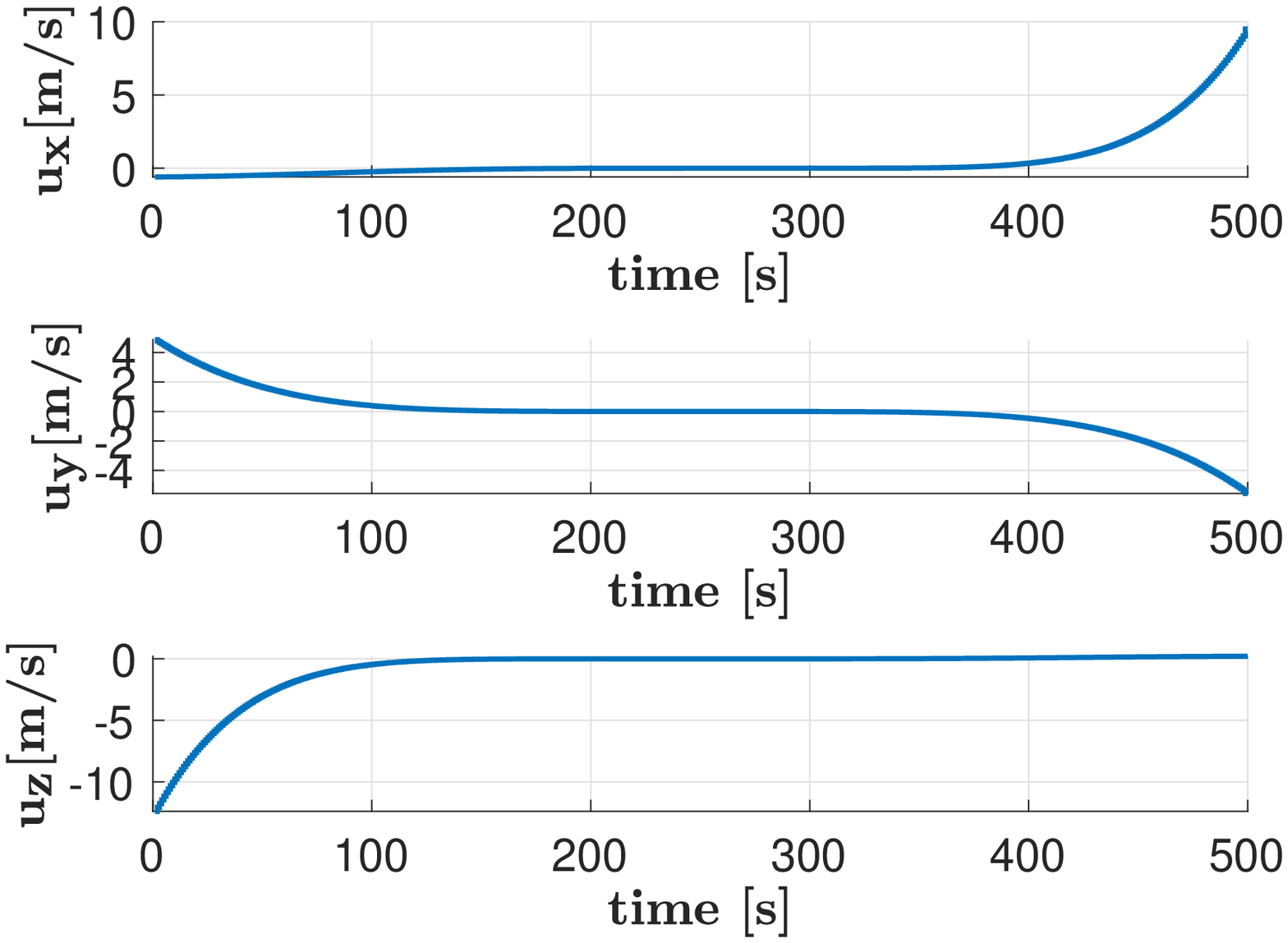}
\caption{$\boldsymbol{u}_\text{koop}$ for far-field rendezvous}
\label{fig:}
\end{subfigure}
\begin{subfigure}{0.41\textwidth}
\includegraphics[scale=0.35]{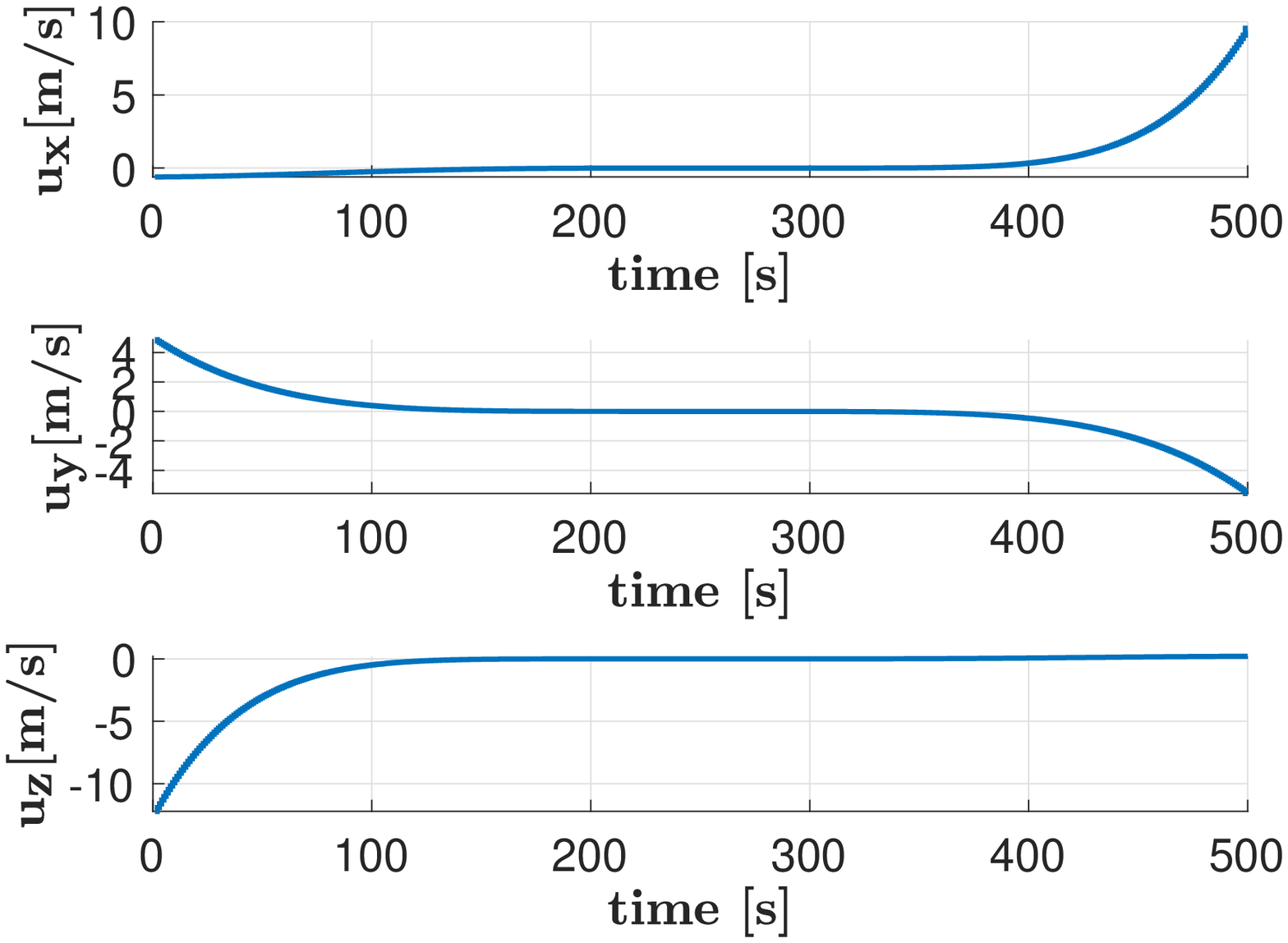}
\caption{$\boldsymbol{u}_{\text{lin}}$ for far-field rendezvous}
\label{fig:}
\end{subfigure}
\caption{Control inputs for far-field rendezvous}
\label{fig:}
\end{figure*}

The main steps of the IRLS algorithm, which will generate control sequences that minimizes the $\ell_2/\ell_1$ control norm given by the performance index in \eqref{eqn:performance_index_nonlinear} are described next.


The value of $ \varepsilon^{[j+1]}$ is now updated to $\min \left\{\varepsilon^{[j]}, \|\boldsymbol{u}_{\text{koop}}^{[j+1]}\|_\infty\right\},$ where $\|\boldsymbol{u}_{\text{koop}}^{[j+1]}\|_\infty$ denotes the $\ell_\infty$-norm of the vector $\boldsymbol{u}_{\text{koop}}^{[j+1]}$. The vector $\boldsymbol{w}^{[j+1]}$ is updated again as follows
\begin{align}
  \boldsymbol{w}^{[j+1]}{(\ell)}=\left(\left(\boldsymbol{u}_{{\text{koop}}}^{[j+1]}(\ell)\right)^{2}+\left(\varepsilon^{[j+1]}\right)^{2}\right)^{-1/4},
\end{align}
for all $\ell\in[1,Nm]$, where $\boldsymbol{u}^{[j+1]}_{{\text{koop}}}(\ell)$ is the $\ell^{\text{th}}$ element of the vector $\boldsymbol{u}_{\text{koop}}^{[j+1]}$ from Eq. \eqref{eqn:control_vector_l2}.
The value of $j$ is now set to $j+1$. Consequently, the updated $\boldsymbol{w}^{[j]}_{(\ell)}$ is used to update the matrix $\mathbf{W}^{[j]}(k)$ and next update matrices the $\boldsymbol{\mathcal{W}}^{[j]}$ and $\boldsymbol{u}_{\text{koop}}^{[j+1]}$ given by Eqns. \eqref{eqn:weight_matirx_big_l2} and \eqref{eqn:control_vector_l2}. This operation is repeated until the control sequence $\boldsymbol{u}_{\text{koop}}$ converges to the optimal control sequence $\boldsymbol{u}^\star_\text{koop}$. If $\varepsilon^{[j]} \notin[0, \bar{\varepsilon}]$, two cases arise. First, if $j<j_{\max },$ then go to Eq. \eqref{eqn:weight_matrix_l2} and if $j=j_{\max }$, then conclude that the algorithm failed to converge. Hence it is suggested to set a larger $j_{\max}$ to increase the chances of success. Else if $j $ is less than or equal to $ j_{\max }$ and $\varepsilon^{[j]} \in[0, \bar{\varepsilon}],$ then Algorithm \ref{algo:irls_sparse_l2/l1} is terminated successfully.
The pseudo code for the IRLS algorithm is given in Algorithm \ref{algo:irls_sparse_l2/l1}.

\section{Numerical simulations}\label{sec:numerical_simulation_results}
Simulation studies presented in this section have been carried out using MATLAB R2020a on Intel Core i7 2.2GHz processor.
Two cases are considered. First, we consider a short-field rendezvous in which the distance between the target spacecraft and active spacecraft is much less than the distance between the planet and the target spacecraft (i.e. $r\ll R$). Second, we consider the case for far-field rendezvous in which $R \approx r$. The target spacecraft is moving in an elliptical orbit whose semimajor axis is equal to $6763\times10^3\mathrm{m}$ and its eccentricity $e=0.73074$.
\begin{table}[H]
\centering
    \centering
    \scalebox{0.93}{
    \begin{tabular}{|c|c|}
\hline Initial state $\boldsymbol{x}_{0}$ & {$(10^3\mathrm{m},\;-10^3\mathrm{m},\;10^3\mathrm{m},\;3\mathrm{m} / \mathrm{s},\;3\mathrm{m} / \mathrm{s},\;-3\mathrm{m} / \mathrm{s})$} \\
\hline Final state $\boldsymbol{x}_{f}$ & {$(0\mathrm{m},\;0\mathrm{m},\;0\mathrm{m},\;0\mathrm{m} / \mathrm{s},\;0\mathrm{m} / \mathrm{s}\;0\mathrm{m} / \mathrm{s})$} \\
\hline
    \end{tabular}
    }
    \caption{Parameters for the short-field spacecraft rendezvous mission}
       \label{tab:mission_parameters_short_field}
    \end{table}
    
    \begin{table}[H]
\centering
    \centering
    \scalebox{0.93}{
    \begin{tabular}{|c|c|}
\hline Initial state $\boldsymbol{x}_{0}$ & {$(10^5\mathrm{m},\;-10^5\mathrm{m},\;10^5\mathrm{m},\;3\mathrm{m}/\mathrm{s},\;3\mathrm{m}/\mathrm{s},\;-3\mathrm{m}/\mathrm{s}) $} \\
\hline Final state $\boldsymbol{x}_{f}$ & {$(0\mathrm{m},\;0\mathrm{m},\;0\mathrm{m},\;0\mathrm{m}/\mathrm{s},\;0\mathrm{m}/\mathrm{s},\;0\mathrm{m}/\mathrm{s})$} \\
\hline
    \end{tabular}
    }
    \caption{Parameters for the far-field spacecraft rendezvous mission}
       \label{tab:mission_parameters_far_field}
    \end{table}

\begin{table}[H]
\centering
\begin{subtable}[t]{0.51\textwidth}
    \centering
    \begin{tabular}{|c|c|c|}
       \hline Terminal state error & Koopman & Linear  \\
\hline Short-field rendezvous & 1.6246 & 4.7369 \\
\hline Far-field rendezvous & 2.9320 & 605.6255\\
\hline
    \end{tabular}
    \end{subtable}
    \caption{$\ell_2$ norm of the terminal state error}
       \label{tab:terminal_state_error}
    \end{table}
The nonlinear dynamics \eqref{eqn:nonlinear_space_rendezvous} is discretized using fourth order Runga Kutta method with discretization step $T$ equal to 1s and $N$ equal to 500. For the Koopman operator we consider $N_k=120$. 
To generate the sequence of data $\boldsymbol{x}(k)$ for $k\in\{0.\dots,d\}$, we sample $1000$ initial conditions which are taken from the uniform distribution over $[-1,1]^6$ . For each sample, we apply control inputs $\boldsymbol{u}(k)$ which are taken randomly from a uniform distribution over $[-1,1]^3$. Then, for each sample of randomly generated initial conditions, we use the discrete nonlinear dynamics in Eq. \eqref{eqn:discrete_nonlinear_equation} to propagate the dynamics with the given control inputs $\boldsymbol{u}(k)$. For each initial condition, we simulate/propagate 2000 states along each trajectory. This data generation process results in matrices $\boldsymbol{X}\;,\boldsymbol{U}$ and $\boldsymbol{Y}$ of size $6\times2\cdot 10^6$. Therefore, the total number of data points is equal to 1000$\times$2000 =$2\cdot10^6$.
The following set of observable functions were used in our simulations:
\begin{align*}
    [g_1\;g_2\;g_3\;g_4\;g_5\;g_6\;]&=[x\;y\;z\;\dot{x}\;\dot{y}\;\dot{z}]\\
    [g_7,g_8,g_9,g_{10},g_{11},g_{12},g_{13}]&=\frac{[1,\dot{x},\dot{y},\dot{z},x,y,z]}{{(1+x^2+y^2+z^2)}^{\frac{3}{2}}}\end{align*}
    \begin{align*}
       [g_{14},g_{15},g_{16}]&=\frac{[x^2\dot{x},y^2\dot{y},z(z-\|\boldsymbol{x}_0\|_2)\dot{z}]}{{(x^2+y^2+(z-\|\boldsymbol{x}_0\|_2)^2)}^{\frac{5}{2}}}\\
         [g_{17},g_{18},g_{19}]&=\frac{[x,y,z]}{{[x^2+y^2+(z-\|\boldsymbol{x}_0\|_2)^2}]^{\frac{3}{2}}}
\end{align*}
where $ g_i(\boldsymbol{x})=1/\sqrt{1+\alpha_i^2}$ ,
$\alpha_i=\sum_{j=1}^6(\boldsymbol{x}(j)^2-\boldsymbol{c}_i(j)^2)$ and $\boldsymbol{c}_i$ is a random vector taken from a uniform distribution over $[-1,1]^6$, for $i\in[20,120]_d$.
\subsection{Short-field rendezvous ($r\ll R$)}
Consider a scenario in which the active spacecraft is performing a short-field rendezvous with a target spacecraft. In this case, $10^3\mathrm{m}\approx r\ll R\approx 10^6\mathrm{m}$. The control inputs $\boldsymbol{u}_{\text{lin}}$ and $\boldsymbol{u}_\text{koop}$ are computed by using the linearized and the lifted space linear dynamics respectively. It is observed from Fig. \ref{fig:states_evolution_short_rendezvous} that these control inputs when applied to the nonlinear discrete rendezvous dynamics in Eq. \eqref{eqn:discrete_nonlinear_equation} can steer the active spacecraft to the desired final states. This is mainly because for the short-field rendezvous, the linearized rendezvous equations can represent the nonlinear spacecraft rendezvous dynamics relatively well.

\subsection{Far-Field rendezvous ($R\approx r$)}
Now, consider the case of far-field rendezvous where $R\approx r\approx 10^6\mathrm{m}$. Again, the control inputs $\boldsymbol{u}_{\text{lin}}$ and $\boldsymbol{u}_\text{koop}$ are generated using the linearized and lifted space dynamics respectively. It is observed that $\boldsymbol{u}_{\text{lin}}$ is not able to steer the active spacecraft to the desired final states as shown in Fig. \ref{fig:states_evolution_far_rendezvous}. However, $\boldsymbol{u}_\text{koop}$ is able to steer the active spacecraft to the desired final states. It can also be observed from Table \ref{tab:terminal_state_error} that the $\ell_2$-norm of the terminal state error is orders of magnitude higher for short-field rendezvous than in the case of far-field rendezvous.


\section{Conclusions}\label{sec:conclusions}
We have presented an iterative scheme for computation of approximate solutions to the minimum-fuel far-field spacecraft rendezvous problem for a thrust vectoring spacecraft. The proposed approach uses Koopman operator to convert the nonlinear dynamics into a approximate higher dimension linear system defined over a lifted state space. The lifted linear state space model is then used together with an Iteratively Recursive Least Squares (IRLS) algorithm to generate approximate solutions for control inputs to minimize the fuel consumption in case of a thrust vectoring spacecraft. Through numerical simulations we showed that for far-field rendezvous, the generated control inputs using the linearized dynamics is not able to steer the active spacecraft from initial to desired final states with good accuracy. By contrast, the control input generated based on lifted space dynamics is able to steer the states for both short-field and far-field rendezvous and is observed to show improved accuracy by orders of magnitude.
\section{Appendix}
In this section, we present the Tschauner–Hempel (T–H) linearized equations for spacecraft rendezvous. Consider the following linearized rendezvous equation given by: 
\begin{align}
    \dot{\boldsymbol{x}}(t)={A}_c(t)\boldsymbol{x}(t)+{B}_c(t)\boldsymbol{u}(t)
    \label{eqn:linearized_state_space_model}
\end{align}
where $A_c(t)$ and $B_c(t)$ are given as follows
\begin{align}
{A}_{c}(t) &=\left[\begin{array}{cc} \mathbb{O}^{3\times3} & \mathbb{I}^3\\ A_1 & A_2 \end{array}\right]
\end{align}
where $A_c\in\mathbb{R}^{6\times 6}$ and $B_c=[\mathbb{O}^{3\times 3}\quad \mathbb{I}^{3}]^\mathrm{T}\in\mathbb{R}^{6\times 3}$ are the state and input matrices respectively. Matrices $A_1$ and $A_2$ are given by:
\begin{align}
& A_1=
    \begin{bmatrix}
    \omega^{2}-k \omega^{\frac{3}{2}} & 0& \dot{\omega}  \\ 0 &-k\omega^{3/2}& 0\\ -\dot{\omega} &0 & \omega^{2}+2 k \omega^{3 / 2}
    \end{bmatrix}
\end{align}
\begin{align}
A_2=
    \begin{bmatrix}
     0 & 0& 2 \omega \\ 0&0&0\\ -2 \omega &0&  0
    \end{bmatrix}.
\end{align}
If $R\gg r$, then the system in \eqref{eqn:nonlinear_space_rendezvous} can be linearized about the origin and can be described by the following non-autonomous discrete-time state space model:
\begin{align}
    \boldsymbol{x}(k+1)=A(k)\boldsymbol{x}(k)+B(k)\boldsymbol{u}(k), \;\; {k\in[0,N-1]_d}
    \label{eqn:linear_system_discrete}
\end{align}
where the matrices $A(k)$ and $B(k)$ are defined as follows:
\begin{subequations}
\begin{align}
    & {A}(k)=\Phi(t_{k+1},t_{k}),\\
    & {B}(k)=\int_{t_{k}}^{t_{k+1}}\Phi(t_{k+1},\sigma){B}_c d\sigma,
    \label{eqn:discrete_linear_state_space_model}
\end{align}
\end{subequations}
where $\Phi$ is the state transition matrix. Using Eq. \eqref{eqn:linear_system_discrete}, it follows that the terminal state at $k=N$ is given by
\begin{align}
     & \boldsymbol{x}(N)=\Phi_d(N,0)\boldsymbol{x}(0)+\sum_{\tau=0}^{N-1}\Phi_d(N,\tau+1)B(\tau), \boldsymbol{u}(\tau).
    \label{eqn:discrete_linear_sytem_final}
\end{align}
where the state transition matrix of the discrete-time system \eqref{eqn:linear_system_discrete}, $\Phi_d(k,m)$, is introduced as follows:
\begin{align}
\Phi_d(k,m) =\left\{ \begin{array}{l}
A(k-1)\dots A(m),  \quad k>m\geq 0,
\\
\mathbf{I}_6, ~~\qquad\qquad\;\;\;\; \qquad k=m,
\end{array}
\right.
\end{align}
where $k$ and $m$ are non negative integers. 
From Eq. \eqref{eqn:discrete_linear_sytem_final}, the terminal state $\boldsymbol{x}(N)=\boldsymbol{x}_f$ can be written in a compact form as follows:
\begin{align}
    \boldsymbol{x}(N)=\boldsymbol{\beta}+\boldsymbol{C}_N\boldsymbol{u},  \quad 
    \label{eqn:terminal_constraint_linear_cn_beta}
\end{align}
where $\boldsymbol{C}_{{N}}$, $\boldsymbol{u}_{\text{lin}}$ and $\boldsymbol{\beta}$ are given by
\begin{subequations}
\begin{align}
    & \boldsymbol{u}_{\text{lin}} =[\boldsymbol{u}(0)^\mathrm{T},\;\boldsymbol{u}(1)^\mathrm{T},\;\dots\boldsymbol{u}(N-1)^\mathrm{T}]^\mathrm{T},\\
    & \boldsymbol{C}_N =[\Phi_d(N,1)B(0),\;\Phi_d(N,2)B(1),\dots B(N-1)],\\
    & \boldsymbol{\beta} =\Phi_d(N,0)\boldsymbol{x}(0).
    \label{eqn:beta_linear}
\end{align}
\end{subequations}
 \bibliography{main.bib}

\end{document}